# Weyl Semimetals: From Principles, Materials To Applications

Mengyuan Zhong[1,2,#], Nam Thanh Trung Vu[1,3,#], Wenhao Zhai[1,3,#], Jian Rui Soh[3,1], Yuanda Liu[1], Jing Wu[1,4], Ady Suwardi[1,5], Huajun Liu[1], Guoqing Chang[6], Kian Ping Loh[7], Weibo Gao[6,*], Cheng-Wei Qiu[8,*], Joel K. W. Yang[9,*] and Zhaogang Dong[1,3,9,*]

[1]Institute of Materials Research and Engineering (IMRE), Agency for Science, Technology and Research (A*STAR), 2 Fusionopolis Way, Innovis #08-03, Singapore 138634, Republic of Singapore

[2]Department of Materials Science and Engineering, National University of Singapore, 9 Engineering Drive 1, 117575, Singapore

[3]Quantum Innovation Centre (Q.InC), Agency for Science Technology and Research (A*STAR), 2 Fusionopolis Way, Innovis #08-03, Singapore 138634, Republic of Singapore

[4]School of Electronic Science & Engineering, Southeast University, Nanjing, 211189, China

[5]Department of Electronic Engineering, The Chinese University of Hong Kong; Sha Tin, New Territories, Hong Kong SAR 999077, Hong Kong

[6]Division of Physics and Applied Physics, School of Physical and Mathematical Sciences, Nanyang Technological University, Singapore 637371, Singapore

[7]Department of Chemistry, National University of Singapore, 3 Science Drive 3, Singapore 117543, Singapore

[8]Department of Electrical and Computer Engineering, National University of Singapore, 4 Engineering Drive 3, Singapore 117576, Singapore

[9]Singapore University of Technology and Design, Singapore 487372, Singapore

[#]These authors equally contribute to this work.

*Correspondence and requests for materials should be addressed to C.-W.Q. (email: chengwei.qiu@nus.edu.sg), W.G. (email: wbgao@ntu.edu.sg), J.K.W.Y. (email: joel_yang@sutd.edu.sg), and Z.D. (email: Zhaogang_dong@sutd.edu.sg).

**Abstract**

Weyl semimetals have attracted significant interest in condensed matter physics and materials science, due to their unique electronic and topological properties. These characteristics not only deepen our understanding of fundamental quantum phenomena, but also make Weyl semimetals promising candidates for advanced applications in electronics, photonics, and spintronics. This review provides a systematic overview of the field, covering theoretical foundations, material synthesis, engineering strategies, and emerging device applications. We first outline the key theoretical principles and distinctive properties of Weyl semimetals, followed by an examination of recent advancements that enhance their functional versatility. Finally, we discuss the critical challenges hindering their practical implementation and explore future development directions, along with the potential for expanding and enhancing their existing range of applications. By integrating discussions of both opportunities and obstacles, this review offers a balanced perspective on current progress and future directions in Weyl semimetal research.

## 1. Introduction

Weyl semimetals are a novel class of material that has attracted significant attention in recent years due to their unique topological and electronic properties. These materials are characterized by a unique band structure, where the conduction band and valence band intersect at discrete points in momentum space, known as Weyl nodes.[1] These nodes appear in pairs, each consisting of two nodes with opposite chirality, and arise only in systems, where at least time-reversal or spatial-inversion symmetry is broken.[2, 3] Meanwhile, a notable outcome of Weyl nodes is their role as monopoles of Berry curvature, giving rise to topologically protected surface states. These surface states are intrinsically robust against local perturbations, including atomic defects and surface adsorption, due to their topological origin.[4] Additionally, Weyl semimetals exhibit a variety of other interesting properties, including pronounced spin-orbit coupling,[5] strong Berry curvature effects,[6] topological Hall responses[7] and high carrier mobility,[8] as well as the ability to undergo phase transitions under external perturbations such as magnetic fields,[9] strain,[10] or chemical doping.[11, 12]

Owing to their unique properties, Weyl semimetals have attracted significant research interest over the past decade, encompassing both fundamental investigations and application-driven studies. In fundamental research, considerable effort has been dedicated to engineering and optimizing ideal Weyl semimetals, characterized by a band structure with a single pair of Weyl nodes. Techniques, such as the application of external magnetic fields, pressure, and element doping, have been employed to achieve these objectives,[13, 14] as will be discussed in detail in subsequent sections. Furthermore, beyond their topological characteristics, the theoretical prediction of superconductivity in Weyl semimetals has garnered substantial attention, motivating experimental efforts to observe this phenomenon.[15, 16] On the application front, active research is underway to harness Weyl semimetals for advanced technologies, including high-performance magnetic sensors,[17] infrared photodetectors,[18, 19] compact optical isolators and circulators,[20] and quantum computing platforms.[21]

In this review, we begin by outlining the fundamental principles that define Weyl semimetals, with a focus on their distinctive topological features and the material systems in which these phenomena have been observed. Next, we discuss their unique physical properties, supported by key experimental findings. We then examine various material engineering techniques employed to modify and enhance these properties, followed by an overview of efforts to explore their potential applications across diverse technological domains. Finally, we address the current challenges in the field and highlight potential breakthroughs that may shape the future of Weyl semimetal research.

## 2. Overview

The concept of Weyl fermions was first theoretically proposed by Hermann Weyl in 1929,[22] as a massless solution to the relativistic Dirac equation in the context of particle physics. Despite their theoretical foundation, these hypothetical massless charge carriers remained elusive until the early 21$^{st}$ century. The breakthrough came with the prediction that certain solid-state materials, known as Weyl semimetals, could satisfy the required symmetry properties to host quasiparticles that mimic the behaviour of Weyl fermions. In 2015, Weyl semimetals were experimentally observed in materials, such as TaAs,[23] NbAs,[24] TaP[25] and related compounds.[26, 27] The existence of Weyl nodes in these materials was confirmed through the observation of Fermi arcs using angle-resolved photoemission spectroscopy (ARPES)[23] and the detection of linear dispersion via scanning tunnelling microscopy/spectroscopy.[28, 29] Depending on their specific dispersion characteristics near the Weyl nodes, these material are further categorized into type I or type II Weyl semimetals.[30] Regardless of type, Weyl semimetals exhibit "massless" charge carriers, as evidenced by their exceptionally high carrier mobility.[17]

Following the discovery of Weyl semimetals, substantial progress has been made in manipulating the reciprocal space location and energy of Weyl nodes to achieve ideal Weyl semimetals.

Researchers have utilized various strategies to modulate the Weyl phase, including elemental doping to alter the crystal structure,[11] thereby inducing band structure crossings. Only recently, an ideal Weyl semimetal has finally been experimentally realized through the precise doping of Cr into $Bi_2Te_3$.[31] Owing to the absence of trivial states near the Fermi level, upon Cr doping, $Bi_2Te_3$ undergoes a smooth transition from topological insulator to Weyl semimetal with a clean electronic structure. Besides elemental doping, other techniques, such as Moiré engineering and external stimuli, including magnetic fields, pressure, and strain, have been employed to fine-tune the electronic properties of these materials.[32, 33] In addition to the identification of Weyl nodes in layered materials, like the $WTe_2$ family, researchers have explored how the topological properties of Weyl semimetals evolve as their dimensionality is reduced, transitioning from bulk three-dimensional systems to thin films, and two-dimensional materials. Various methods have been developed to fabricate Weyl semi-metallic thin films, including molecular beam epitaxy (MBE),[34] chemical vapor deposition (CVD),[35, 36] chemical vapor transport,[37] temperature gradient growth method,[26, 27] solid-state reaction [38] and mechanical exfoliation. Table 1 summarizes different Weyl semimetals along with their corresponding preparation methods.

**Table 1. Preparation methods of Weyl semimetals**

| Type | Weyl semimetals | Preparation method | Reference |
|---|---|---|---|
| Type I | $(NbSe_4)_2I/(TaSe_4)_2I$ | Temperature gradient growth | 26, 27 |
| | NbAs | CVT, CVD | 35, 36 |
| | TaAs | CVT | 35, 36, 39 |
| | TaP | CVT | 37 |
| | $PtBi_2$ | Self-flux method | 38 |
| | NbP | Solid-state reaction | 39 |
| | $K_2Mn_3(AsO_4)_3$ | Solid-state reaction | 40 |
| | $Co_2MnGa$ | MBE | 41 |
| | $Co_3Sn_2S_2$ | Heat treatment | 42-45 |
| Type II | NdAlSi | Self-flux method | 2 |
| | $MoTe_2$ | CVT | 8 |
| | $Mo_{1-x}Ta_xTe_2$ | Self-flux method | 15 |
| | $Mn_3Sn/Mn_3Ge$ | Magnetron sputtering, Bridgman method | 46, 47 |
| | RhSi | Vertical Bridgman method | 48 |
| | $WTe_2$ | MBE, CVD, Heat treatment | 49-51 |
| | $TaIrTe_4$ | Te flux solid-state reaction | 52 |
| | $NbIrTe_4$ | Solid-state reaction, mechanical exfoliation | 53 |
| | β-$WP_2$ | CVT | 54 |
| | CeAlGe | Self-flux method | 55 |
| Topological materials potentially exhibiting Weyl semi-metallic phase | $EuCd_2As_2$ | NaCl/KCl flux growth | 9 |
| | CeAlSi | Self-flux method | 12 |
| | $Cd_3As_2$ | MBE | 9, 13 |
| | $Bi_{0.96}Sb_{0.04}$ | MBE | 56 |
| | MCl (M=Sc, Y, La) | CVD, mechanical exfoliation | 57 |
| | SmAlSi (Sm=Ce, Pr, Nd) | Self-flux method | 12, 58 |
| | TbPtBi | Self-flux method | 59 |
| | CoSi Film | Flash-Lamp Annealing | 60 |

Beyond fundamental studies, recent advancements have increasingly explored the diverse potential applications of Weyl semimetals. Their exceptionally high carrier mobility positions them as promising candidates for advanced thermoelectric applications, where achieving high electrical conductivity alongside low thermal conductivity is critical.[61] Additionally, the spin-polarized, topologically protected surface states of Weyl semimetals offer unique opportunities for innovative spin-based devices.[62] Furthermore, their remarkable nonlinear optical properties open new avenues in photonics, including the development of advanced optical modulators and detectors.[63] Finally, Weyl semimetals are being explored for photocatalytic applications, with their robust electronic states being utilized to enhance light-driven chemical reactions.[64]

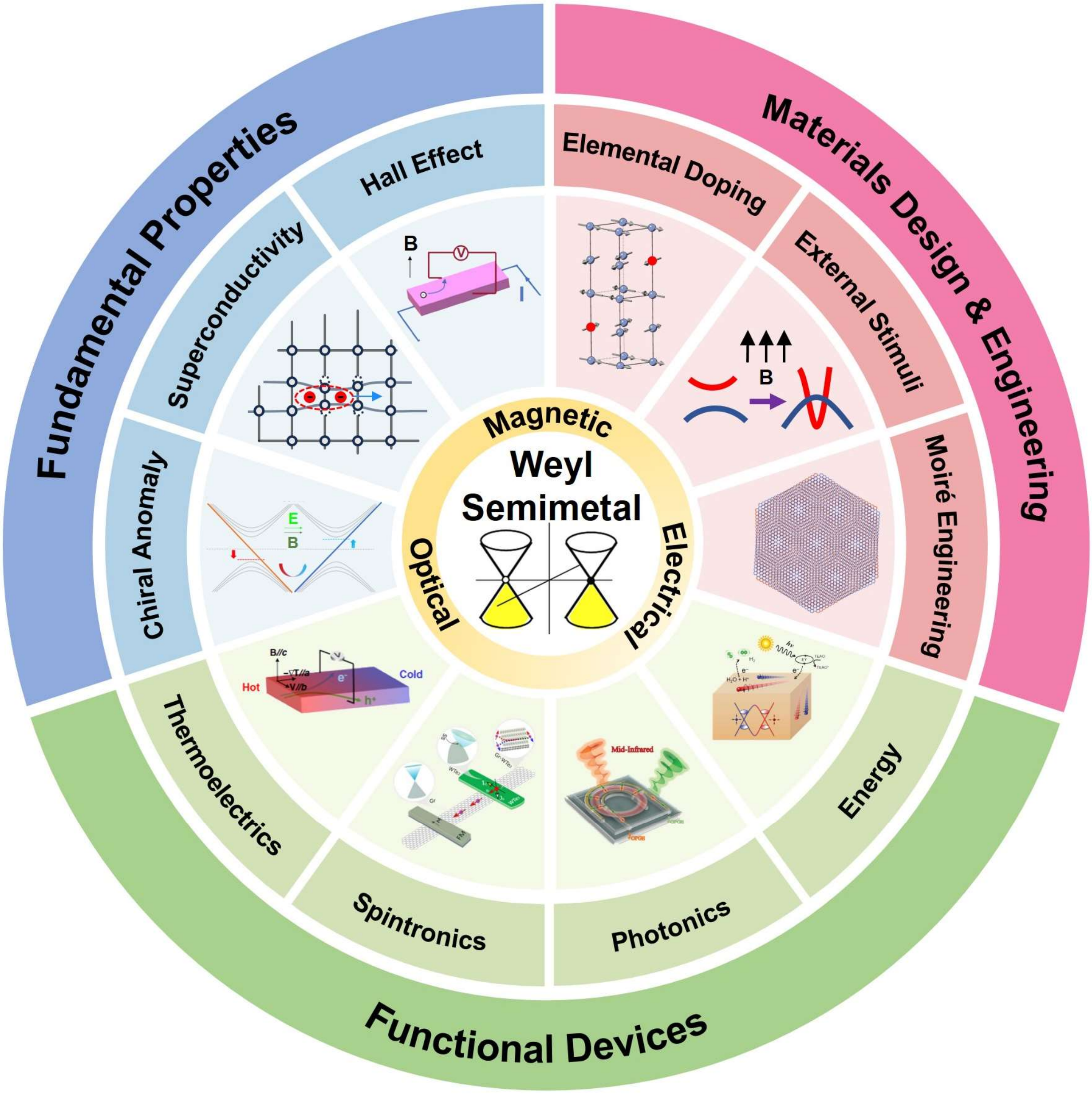

**Figure 1. Weyl semimetals in various research fields.** This figure illustrates the three primary research focus in Weyl semimetals: Fundamental properties, materials design and engineering, and functional devices. The *fundamental properties* section covers the study of quantum phenomena, such as chiral anomaly, superconductivity and Hall effect. The *materials design and engineering* section highlights efforts to realize and optimize Weyl semimetals through methods, such as elemental doping, external stimuli, and Moiré engineering. Finally, the *functional devices* section explores potential applications of Weyl semimetals in various areas, such as thermoelectrics,

spintronics, photonics, and energy. Together, these categories represent the interconnected framework of both theoretical and experimental advancements in the field.

## 3. Theoretical Background

### 3.1. Weyl semimetals and their analogy in high energy physics

In fact, Weyl semimetals derive their name from high energy physics, specifically from Hermann Weyl (1885–1955), who proposed the Weyl equation in 1929 as a massless version of the Dirac equation for spin $\frac{1}{2}$ particles[65]. The equation is expressed as:

$$i\hbar\frac{\partial\Psi}{\partial t} = \widehat{H}\Psi = \pm c\mathbf{p}\cdot\boldsymbol{\sigma}\Psi. \qquad (1)$$

Here, $\hbar$ is the reduced Planck constant, c is the speed of light, $\mathbf{p} = (p_x, p_y, p_z)$ is the momentum operator, $\boldsymbol{\sigma} = (\sigma_x,\sigma_y,\sigma_z)$ represents the Pauli matrices, and $\psi$ is a two-component Weyl spinor.[32] The solutions of the Weyl equation describes massless charged particles with a linear energy-momentum dispersion, known as Weyl fermions. The ± sign denotes the two chiralities of Weyl fermions: right-handed (+) and left-handed (−).

In condensed matter physics, materials with broken spatial-inversion symmetry, time-reversal symmetry, or both can exhibit electronic bands that intersect at discrete points in reciprocal space. Near these intersection points, the electronic bands display a linear energy-momentum dispersion, analogous to the behaviour of the Weyl fermions described earlier. Consequently, these points are termed Weyl nodes or Weyl points.[66] Importantly, when Weyl nodes are located near the Fermi energy, the material can host quasiparticle excitations that closely mimic the massless, chiral behaviour of Weyl fermions. This unique property enables Weyl semimetals to exhibit exceptionally high carrier mobility, as Weyl electrons are effectively "massless" and are resistant to backscattering processes that preserve their chirality.[67]

Weyl nodes naturally occur in pairs, with each node having opposite chirality, either left-handed or right-handed.[68] However, in material systems where both time-reversal and spatial-inversion symmetry are preserved, these Weyl node pairs merge into a single fourfold-degenerate Dirac point. This degeneracy encompasses both spin states (up and down) as well as opposite momentum directions.[69] Materials exhibiting these characteristics are classified as Dirac semimetals, with prominent examples, including $Cd_3As_2$[70] and $Na_3Bi$.[71] Like Weyl semimetals, Dirac semimetals feature "massless" electrons due to their linear energy-momentum dispersion. However, they lack chirality-related effects, as the contributions from the two opposite-chirality Weyl points cancel each other out. Moreover, the (massless) Dirac points are not protected by topology and can sometimes gap out to give rise to avoided crossings (massive) (Figure 2a). Notably, Dirac semimetals can transition into Weyl semimetals when external perturbations, such as low temperatures, applied magnetic fields, or chemical doping, break either time-reversal or spatial-inversion symmetry.[72-74]

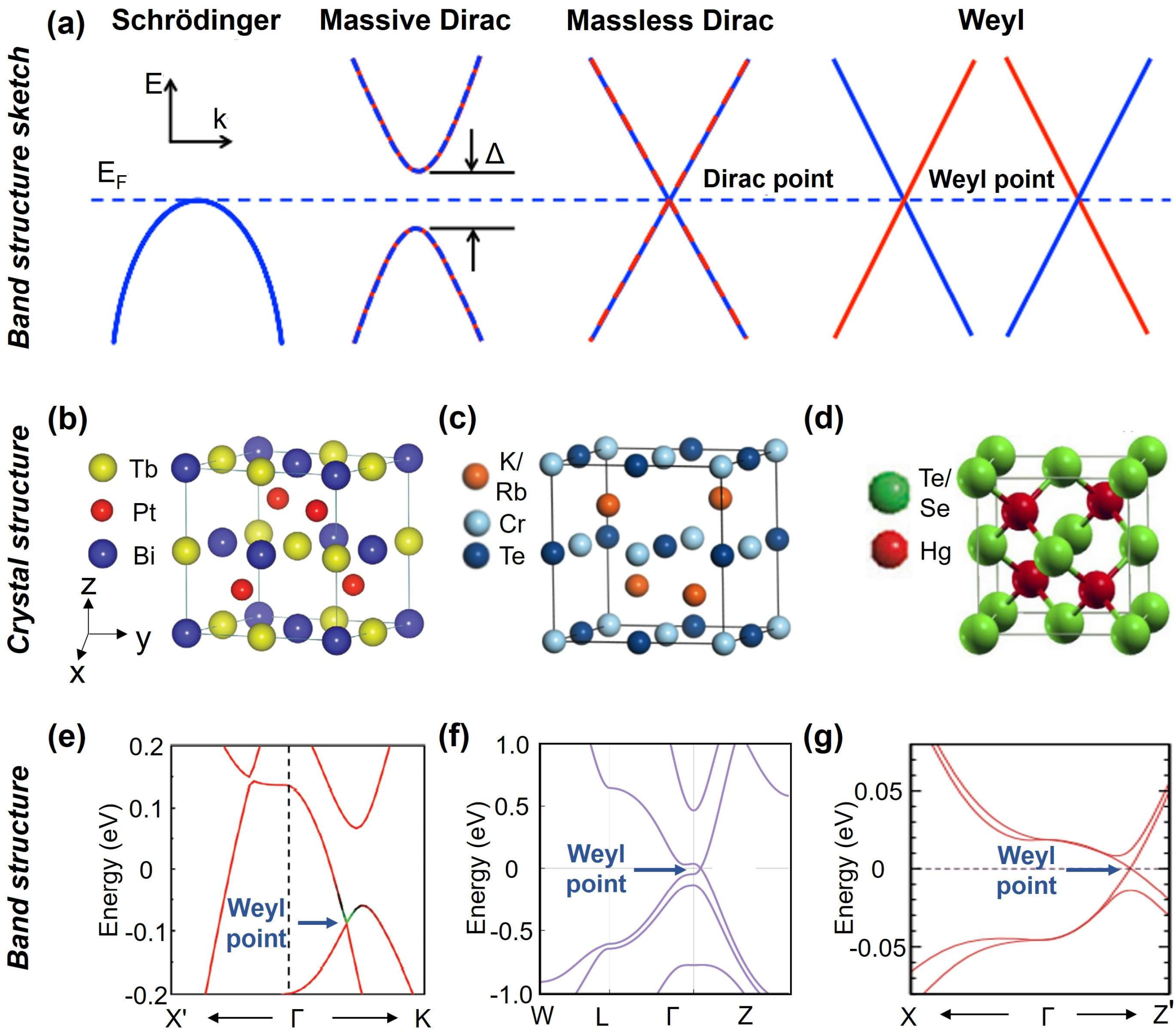

**Fig. 2. Crystal structures and band structures of typical Weyl semimetals.** (a) Schematic of the band structures for Schrödinger, massive Dirac, massless Dirac, and Weyl fermions. Doubly degenerate bands are distinguished by curves with mixed colors, while nondegenerate bands are indicated by uniform colors. (b-d) The crystal structure of (b) TbPtBi; (c) half-Heusler compounds XCrTe (X=K, Rb) and (d) HgTe/HgSe. (e-g) Weyl points in (e) TbPtBi, (f) half-Heusler compounds XCrTe (X=K, Rb) and (g) HgTe/HgSe. Figures are reproduced with permission from (a) Ref.[69], Copyright 2021 American Physical Society; (b, e) Ref.[59], Copyright 2022 Wiley; (c, f) Ref.[75], Copyright 2024 American Physical Society; (d, g) Ref.[10], Copyright 2016 Springer, under a Creative Commons Attribution 4.0 International License (CC BY 4.0).

### 3.2. Examples of Weyl semimetals and their corresponding band structure

Figures 2b and 2e illustrate the crystal structure and band structure of TbPtBi, respectively. In Firure 2b, TbPtBi exhibits a cubic crystal structure with the space group F-43m, where Tb atoms display type-II antiferromagnetic order along the [111] direction. DFT calculations reveal that TbPtBi is a Weyl semimetal, with Weyl points located near the Γ-point, as shown in Figure 2e. These Weyl points arise from band inversion driven by spin-orbit coupling and the magnetic moment of Tb.[59]

Similarly, the Weyl points in XCrTe (X=K, Rb) arise from the magnetic properties of the material, which breaks time-reversal symmetry (Figures 2c and 2f).[75] As a result, the position of the Weyl point can be tuned by rotating the magnetization vector.[44] In these materials, the Fermi surface consists of a single pair of Weyl points, representing an ideal Weyl semimetal state. The clean Fermi surface, characterized by its dominance of Weyl points with minimal contributions from trivial states, makes these materials an excellent platform for exploring chiral Weyl fermion physics and its interaction with spintronics.

Meanwhile, the Weyl points in HgTe-like materials arise from their non-centrosymmetric crystal structure, which breaks spatial-inversion symmetry (Figures 2d and 2g), in combination with the exhibited band inversion.[10] Notably, all Weyl nodes in these materials are precisely located at the Fermi level, without any coexisting trivial states, similar to graphene.

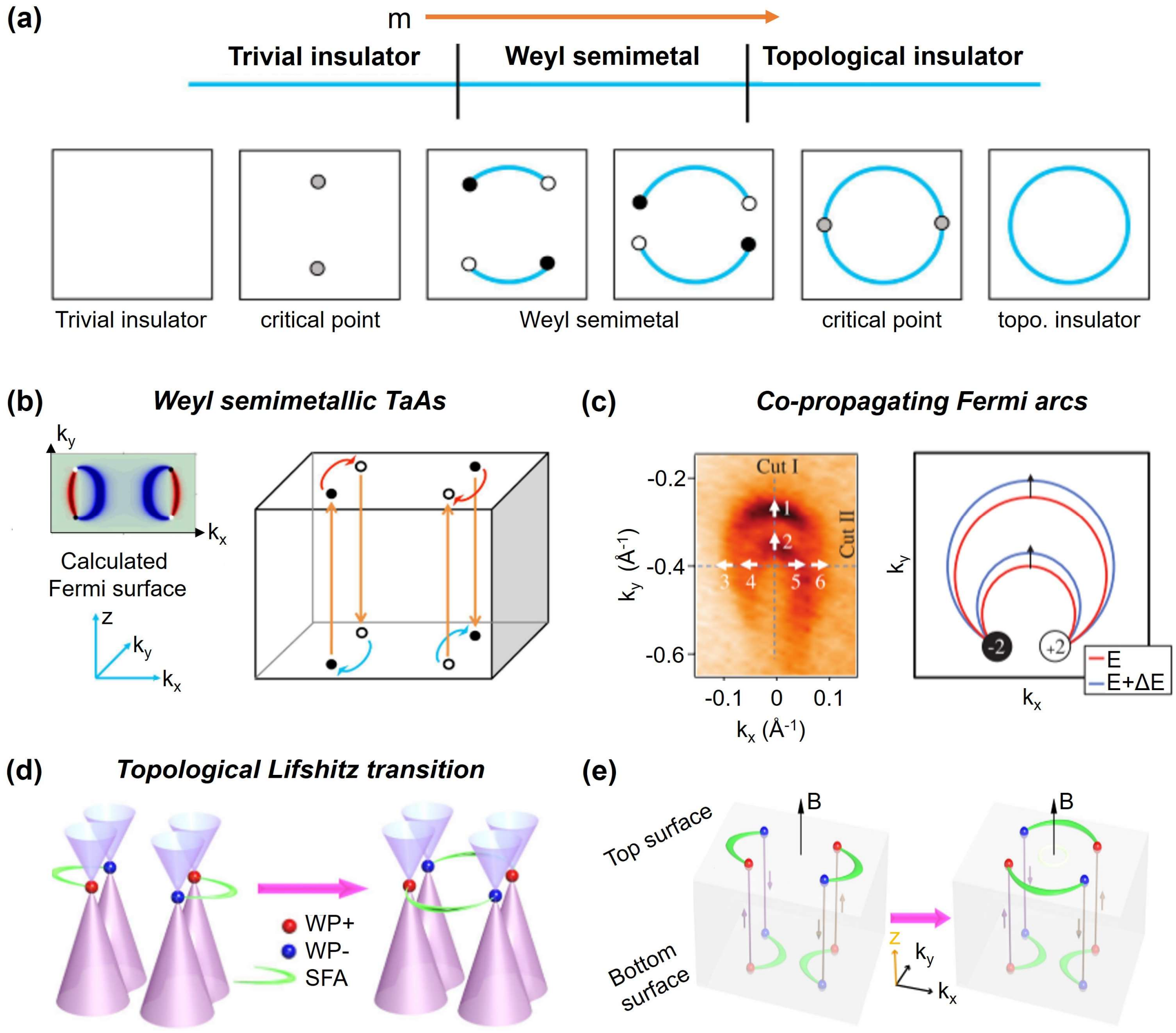

**Figure 3**. **Topological features of Weyl semimetals.** (a) A Weyl semimetal represents an intermediate phase between a trivial insulator and a topological insulator, governed by a tuning parameter *m*. At critical *m*, bulk conduction and valence bands touch, giving rise to Weyl points. An increase in *m* leads to the splitting of Weyl points into pairs with Fermi arcs connecting them. As *m* further increases, the Weyl points annihilate, reopening a bulk gap and transitioning to a topological insulator with gapless surface states. (b) The calculated Fermi surface of TaAs near the X point highlights its topological features. (c) High-resolution ARPES mapping reveals crescent-shaped Fermi arcs, their evolution across binding energies, and the distinction between arcs and closed contours. (d) A schematic depicts the Lifshitz transition of surface Fermi arcs in topological Weyl semimetals. (e) Weyl orbits, weaving chiral bulk states and surface Fermi arcs across top and bottom surfaces under magnetic fields, modify quantum phenomena and physical responses during the Lifshitz transition. Figures are reproduced with permission from (a, b) Ref.[76], Copyright 2015 Springer Nature, under a Creative Commons Attribution 4.0 International License (CC BY 4.0); (c) Ref.[77], Copyright 2015 AAAS; (d, e) Ref.[78], Copyright 2019 Springer Nature, under a Creative Commons Attribution 4.0 International License (CC BY 4.0).

### 3.3. Topological features of Weyl semimetals

An intriguing property of Weyl nodes is their topological nature, meaning they arise from the global properties of the system, such as its symmetry.[32] This topological origin imparts exceptional robustness to Weyl nodes, and consequently to the properties of Weyl semimetals, against local chemical imperfections and perturbations, including atomic defects and surface adsorption. Another distinctive feature of Weyl semimetals is the presence of open Fermi arcs on their surfaces, which connect Weyl nodes of opposite chirality, giving rise to topological surface states. These arcs demonstrate chiral selectivity by linking Weyl nodes with opposite handedness.[79] Unlike the closed Fermi surfaces found in other topological materials, such as topological insulators, the open Fermi arcs are a defining characteristic of Weyl semimetals, setting them apart in the realm of topological materials.[76]

The phase transition between a trivial insulator, Weyl semimetal, and topological insulator can be understood through the evolution of Weyl points and surface states in momentum space (Figure 3a). This process can be visualized by defining a tuning parameter named $m$. In the trivial insulator phase, the bulk bands are fully gapped, with no surface states present. As $m$ increases to a critical value, the conduction and valence bands touch, giving rise to Weyl points. A further increase in $m$ leads to the splitting of these points into pairs of opposite chirality, connected by surface Fermi arcs, marking the Weyl semimetal phase. With a continued increase in $m$, the Weyl nodes eventually annihilate, reopening a full band gap and transforming the material into a topological insulator, being characterized by gapless surface states.

The electronic structure of Weyl semimetals can be effectively studied through ARPES. For instance, ARPES measurements of TaAs near the X-point reveal crescent-shaped Fermi arcs on its Fermi surface (Figure 3b), with their evolution across binding energy mapped in high resolution (Figure 3c).[77] Constant-energy contours illustrate how Fermi arcs connect Weyl nodes and transition into closed contours at specific energy levels. When subjected to a magnetic field, Weyl semimetals undergo what is known as a topological Lifshitz transition,[78] which alters the connectivity of surface Fermi arcs (Figure 3d). Under these conditions, Weyl semimetals also exhibit Weyl orbits, where chiral bulk states are intertwined with surface Fermi arcs connecting opposite surfaces (Figure 3e). The Lifshitz transition, combined with the emergence of Weyl orbits, leads to significant modifications in the quantum transport properties and other physical responses of Weyl semimetals, highlighting the complex interplay between their bulk and surface states.[78]

## 4. Fundamental Properties

Owing to their aforementioned unique characteristics, Weyl semimetals exhibit a range of intriguing quantum phenomena. Among these, the most extensively studied are chiral anomaly,[1] negative magnetoresistance,[80] superconductivity[81], exotic Hall effect[21] and ultrahigh transport lifetime[82-84]. These phenomena have attracted significant research interest due to their potential implications for both fundamental physics and practical applications. The following sub-sections will provide a more detailed discussion of these unique properties.

### 4.1. Chiral anomaly & negative magnetoresistance

The chiral anomaly and negative magnetoresistance are hallmarks of Weyl semimetals, arising from their topological nature and the presence of Weyl nodes in their electronic structure. The chiral anomaly refers to the violation of chiral symmetry, which is the independent conservation of left-handed and right-handed fermions.[85] This phenomenon, which is intrinsically tied to the relativistic behaviour of Weyl fermions, manifests in solid-state systems through Weyl electrons. Various experimental techniques have been employed to probe the chiral anomaly, including nonlocal transport measurements,[21] thermal power,[86] conductivity analyses,[87] and optical pump-probe responses.[88]

Among these characterization approaches, longitudinal magnetoresistance (LMR) experiments are particularly prevalent. Here, LMR refers to the change in electrical resistance of a material when a magnetic field is applied parallel to the electric current. In conventional materials, a positive magnetoresistance is observed, where electrical resistance typically increases under a magnetic field due to the Lorentz force that induces cyclotron motion and enhances scattering. However, in Weyl semimetals, researchers observe negative magnetoresistance, where resistance decreases as the magnetic field strength increases, providing evidence of the chiral anomaly. [89, 90] Specifically, the non-conservation of chiral charge in Weyl semimetals leads to charge pumping between Weyl nodes of opposite chirality, creating an imbalance between left- and right-handed electrons. This process effectively accelerates charge carriers along the direction of the applied electric field, reduces scattering and enhances electrical conductivity.[80] The sharp drop in resistance, along with its dependence on magnetic field strength and orientation, provides strong evidence for the chiral anomaly and the associated existence of Weyl points.

It is important to note that the detection of chiral anomaly through LMR measurements can be influenced by the current injection effect, which occurs whenever a magnetic field is applied. This effect causes the current density to concentrate at the center of the sample, forming a narrow beam, while the edges experience reduced current flow. This uneven distribution results from the suppression of carrier movement in the transverse direction by the applied magnetic field, while carriers are relatively enhanced in the longitudinal direction. Such a nonuniform distribution of current can complicate the interpretation of LMR data, potentially obscuring the intrinsic chiral anomaly effects.[85] Therefore, careful experimental design and theoretical modeling are essential to accurately isolate and confirm the chiral anomaly in Weyl semimetals.

### 4.2. Superconductivity

Although not intrinsically linked to the unique topological properties of Weyl nodes, superconductivity has been observed in Weyl semimetals, positioning them as promising candidates for quantum technology applications, particularly in superconducting topological qubits.[91] This phenomenon was first reported in $MoTe_2$, which undergoes a superconducting transition at 0.10 Kelvin (K) under ambient pressure.[81] Notably, the superconducting transition temperature ($T_c$) increases significantly under high pressure, reaching a maximum of 8.2 K at 11.7 GPa. This pressure-dependent enhancement of $T_c$ follows a dome-shaped phase diagram, illustrating the intricate interplay between superconductivity, structural phase transitions, and electronic topological properties. The pronounced tunability of $T_c$ in Weyl semimetals under external pressure underscores their potential for phase-transition-driven modifications of electronic and quantum states, further broadening their utility in advanced applications.

The superconducting properties of $MoTe_2$ have been further explored using point-contact spectroscopy, which revealed distinct surface superconductivity at ambient pressure.[81, 92] Interfaces between $MoTe_2$ and metals, such as Ag or Cu, demonstrated Andreev reflection, confirming the presence of superconducting order. The average superconducting gap was measured as $0.650 \pm 0.075$ meV, exceeding the conventional BCS value and suggesting nodeless superconductivity potentially influenced by strong-coupling effects. Additionally, "gapless-like" dV/dI curves observed in point contacts indicate the possibility of a topological superconducting state at the surface of $MoTe_2$. This observation links the superconducting behavior to the topologically protected surface state characteristic of Weyl semimetals, highlighting the interplay between topology and superconductivity.[92]

Advancements in the superconducting properties of $MoTe_2$ have been achieved through Ta-doping, which enhances superconductivity and increases the upper critical field.[15] As depicted in Figure 4a, a sharp drop in resistivity is observed in $Mo_{1-x}Ta_xTe_2$ samples at low temperatures, marking the onset of superconductivity for Ta concentrations $x \geq 0.08$. The structural and superconducting phase diagram shown in Figure 4b highlights the relationship between structural phase transitions and superconductivity. With increasing Ta concentration, the structural transition temperature $T_s$ decreases, while the superconducting transition temperature $T_c$ increases, peaking at 7.5 K for x=0.15.[15] These findings underscore the tunable nature of $MoTe_2$'s superconducting and electronic properties, providing a robust platform for exploring topological superconductivity and its potential links to the unique surface states of Weyl semimetals.

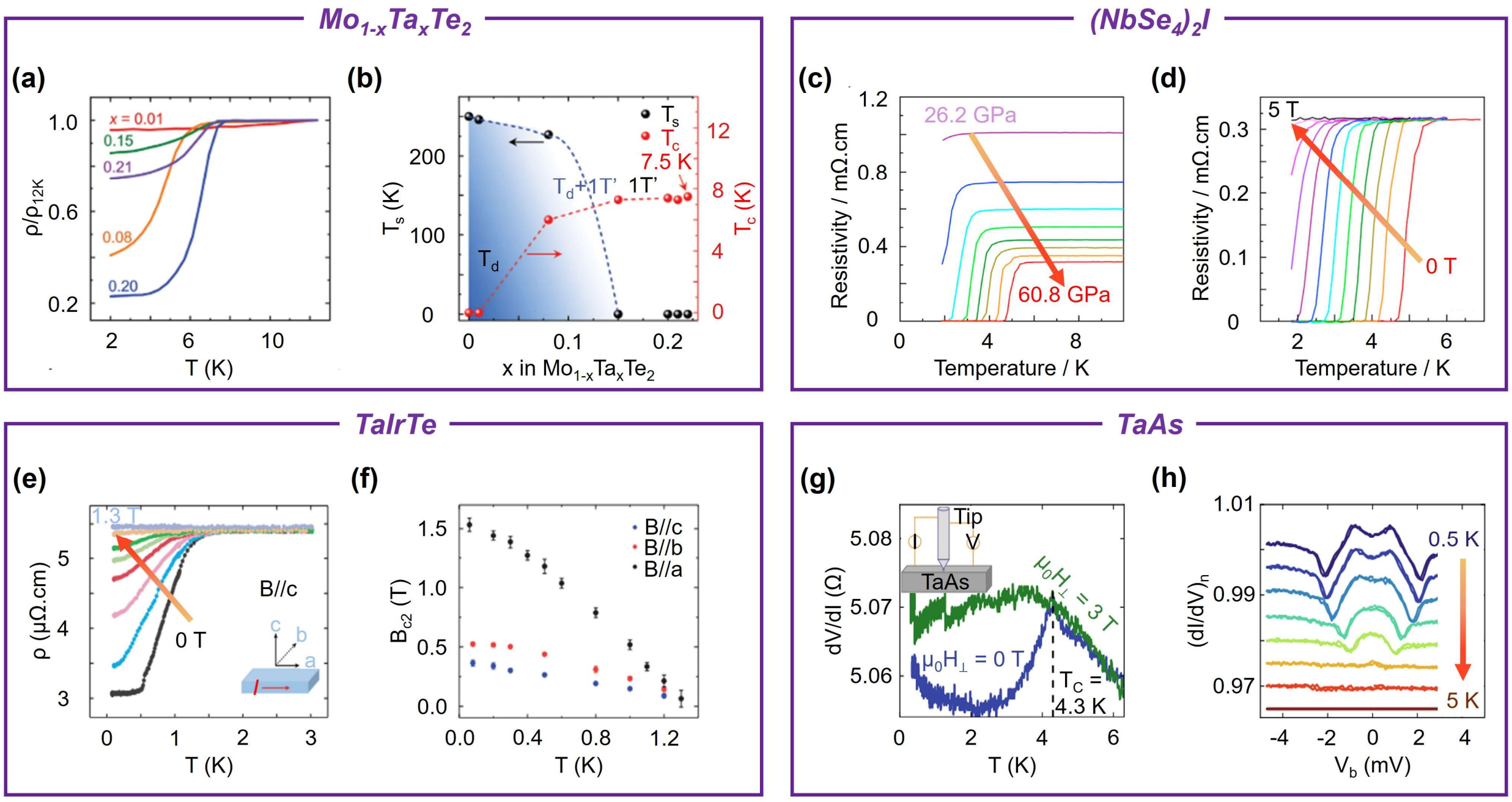

**Figure 4. Superconductivity of Weyl semimetals.** (a) Emergence of superconductivity in $Mo_{1-x}Ta_xTe_2$ at low temperatures for Ta concentrations $x \geq 0.08$. (b) Structural and superconducting phase diagram of $Mo_{1-x}Ta_xTe_2$, highlighting the rise of $T_c$ with increasing Ta content. (c) Pressure-induced superconductivity in $(NbSe_4)_2I$, showing a transition to superconductivity at 36.4 GPa. (d) Suppression of superconductivity in $(NbSe_4)_2I$ under an external magnetic field of 5 T. (e-f) Anisotropic superconductivity in $TaIrTe_4$, with differing critical magnetic fields ($B_{c2}$) along crystal axes. (g-h) Superconducting transition in TaAs point-contact spectra, suggesting unconventional superconductivity induced by the ferromagnetic proximity effect. Figures are reproduced with permission from (a-b) Ref.[15], Copyright 2023 Wiley; (c-d) Ref.[26], Copyright 2021 Elsevier; (e-f) Ref.[16], Copyright 2020 Oxford Academic, under the terms of the Creative Commons CC BY license; (g-h) Ref.[93], Copyright 2020 Elsevier.

In addition to $MoTe_2$, other Weyl semimetals have also been found to exhibit superconducting properties. Using various techniques, researchers have demonstrated the superconductivity of Weyl semimetals, such as $(NbSe_4)_2I$, $TaIrTe_4$ and TaAs. The resistivity of $(NbSe_4)_2I$, as shown in Figure 4c, decreases with the increasing pressure, where it is transitioned to a superconducting state with a critical temperature $T_c$ of 5.2 K at 60.8 GPa.[26] This superconductivity is suppressed under an external magnetic field (see Figure 4d), and it vanishes entirely above 1.8 K under a 5 T field. The anisotropic superconducting behavior of $TaIrTe_4$ (Figures 4e and 4f ) reveals that the critical magnetic field $B_{c2}$ varies significantly along different crystal axes, with B//a showing a much higher $B_{c2}$ (~ 1.5 T) compared to B//c (~ 0.5 T).[16]

Finally, TaAs, one of the first demonstrated Weyl semimetals, can also be induced to superconduct under certain conditions. Wang et al. induced superconducting states in a Weyl semimetal TaAs single crystal by applying a hard point contact using a ferromagnetic tip (e.g. nickel or cobalt).[93] As shown in Figure 4g, the resistance of a ferromagnetic tip-induced superconducting state is plotted as a function of temperature in a Ni/TaAs point contact experiment. At 4.3 K, the resistance drop at zero field indicates a superconducting transition, corresponding to the critical temperature $T_c$. When a 3 T perpendicular magnetic field is applied, the superconducting transition disappears, demonstrating that the superconducting state is suppressed by the magnetic field. In Figure 4h, the normalized point contact spectra at zero magnetic field are displayed for different temperatures. The figure reveals an enhancement in conductance at low bias voltages, which diminishes as the temperature increases, further confirming the presence of a superconducting state. Additionally, characteristic features of a p-wave superconducting state, such as double conductance peaks and valleys, are observed. These features strongly suggest that the induced superconducting states may be unconventional p-wave superconductors.

In summary, these investigations conducted under high pressure and external magnetic fields have provided critical insights into the superconducting behavior of Weyl semimetals. These studies have confirmed the bulk nature of superconductivity through magnetic screening effects and correlated it with the resistivity drop being observed during the superconducting transition. The observed surface superconductivity, pressure-tunable behavior, and potential topological superconducting states underscore the unique properties of Weyl semimetals. These features position Weyl semimetals as pivotal materials for exploring quantum phenomena and advancing their applications in quantum technologies. However, it is important to emphasize that achieving the topological superconductivity required for quantum technologies necessitates verifying whether the observed superconductivity specifically arises from Weyl electrons. This distinction is crucial because, although other trivial electronic states at the Fermi level in the studied non-ideal Weyl semimetals may also contribute to superconductivity, they do not possess the topological characteristics inherent to Weyl electrons. Therefore, unambiguously attributing superconductivity to Weyl electrons is a crucial step towards harnessing their topological properties for practical quantum devices.

### 4.3. Hall effect

Hall measurements are an essential tool for investigating the transport properties of materials, often uncovering complex behaviors in systems with intricate electronic structures. In addition to the ordinary Hall effect, which results from the deflection of charge carriers by the Lorentz force under an external magnetic field, Weyl semimetals exhibit a range of exotic Hall effects due to their topological properties and the presence of Weyl nodes in their electronic band structure. These Weyl nodes act as sources and sinks of Berry curvature, a concept that can be understood as the counterpart of a magnetic field in momentum space and play a central role in driving these unconventional Hall effects. In Figure 5a, left panel illustrates the standard experimental setup for Hall effect measurements, while right panel shows a more delicate angular dependence Hall measurement. Unlike the ordinary Hall effect, which requires an external magnetic field applied perpendicular to the current, anomalous Hall effect (AHE) and nonlinear Hall effect (NLHE) in Weyl semimetals does not necessitate an external magnetic field. This is because the AHE arises from the material's internal magnetization and spin-orbit coupling, while the NLHE is driven by the material's intrinsic Berry curvature.[51, 94]

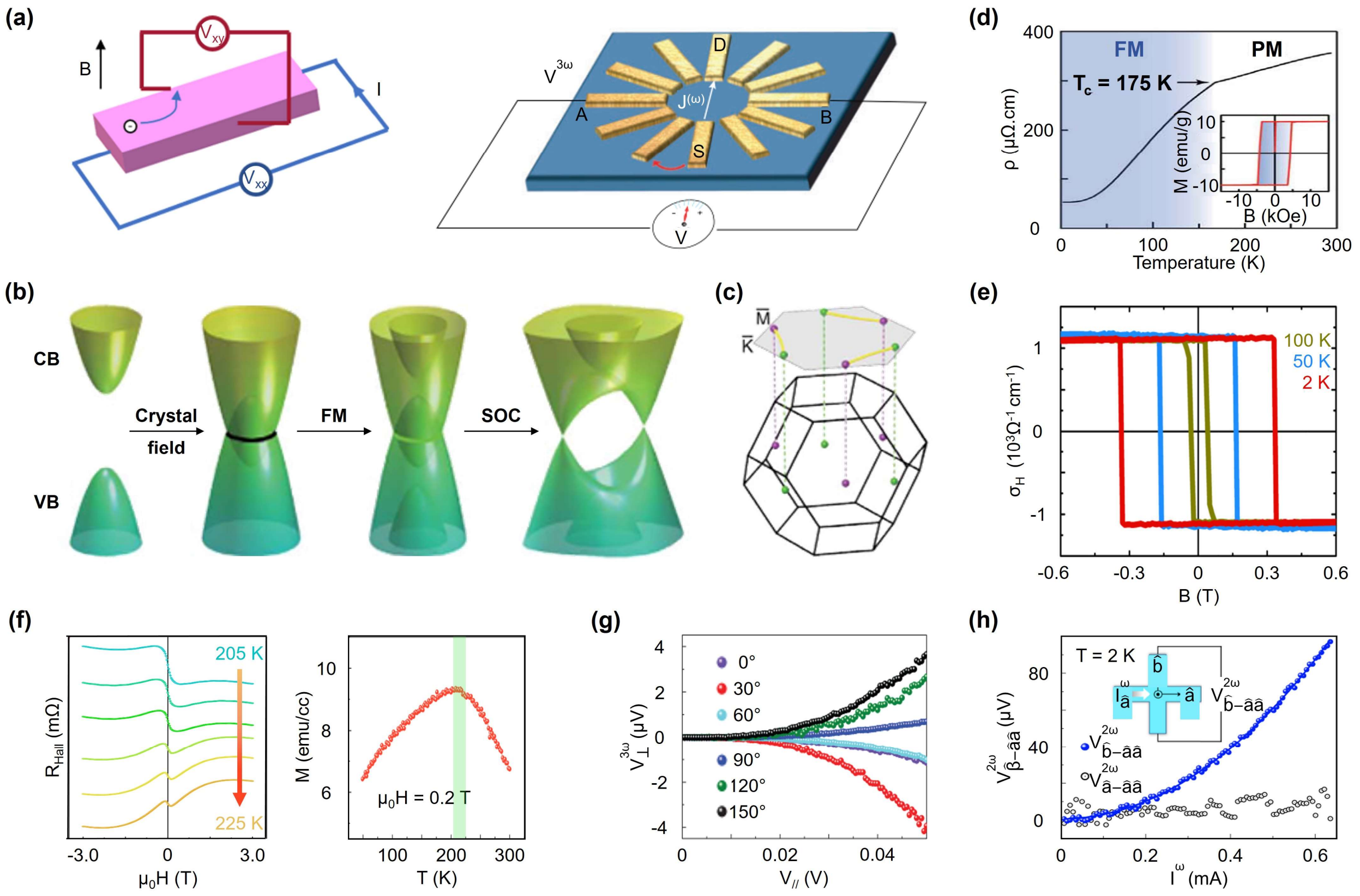

**Figure 5. Various Hall effects in Weyl semimetals.** (a) Schematic of standard Hall effect measurement (left panel) and angular dependence Hall measurement (right panel). The driving electric field is applied across one pair of opposing electrodes, while the transverse voltage is measured across the other pair. By rotating the measurement frame, signals corresponding to different lattice orientations can be recorded. (b) Mechanism of the magnetic Weyl semimetal phase in $Co_3Sn_2S_2$. CB: Conduction band, VB: valence band, FM: ferromagnetism, SOC: spin-orbit coupling. (c) Schematic representation of the bulk and surface Brillouin zones. Depicted along the (001) surface of $Co_3Sn_2S_2$, with Weyl points marked and connected by surface Fermi arcs as yellow line segments. (d) Temperature dependence of longitudinal electrical resistivity, showing a ferromagnetic transition at $T_c$ = 175 K. FM: ferromagnetic phase, PM: paramagnetic phase. (e) Field dependence of Hall conductivity ($\sigma_H$) at 100, 50, and 2 K of $Co_3Sn_2S_2$. (f) Topological Hall effect (left panel) and corresponding magnetization (right panel) in thin films of $Mn_3Sn$ integrated on Si. (g) The nonlinear relationship between the third harmonic transverse voltage $V_{\perp}^{3\omega}$ and the first harmonic longitudinal voltage $V_{\parallel}$, for different lattice orientations in $TaIrTe_4$. (h) Second-order NLHE in $T_d$-$TaIrTe_4$, where the transverse voltage $V_{\hat{b}-\hat{a}\hat{a}}^{2\omega}$ is proportional to the square of the current $I^{\omega}$. Figures are reproduced with permission from (b-d) Ref.[44] , Copyright 2019 AAAS; (e) Ref.[95], Copyright 2018 Springer Nature; (f) Ref.[96], adapted with permission from X. Wang, X. Zhou, H. Yan, P. Qin, H. Chen, Z. Meng, Z. Feng, L. Liu, Z. Liu, Topological Hall effect in thin films of an antiferromagnetic Weyl semimetal integrated on Si, ACS Applied Materials & Interfaces 2023, 15, 7572, Copyright 2023 American Chemical Society; (g) Ref.[97], Copyright 2022 Oxford academic, under the terms of the Creative Commons CC BY license; (h) Ref.[94], Copyright 2021 Springer Nature.

One of the most extensively studied phenomena in Weyl semimetals is AHE, which occurs in the absence of an external magnetic field. In conventional ferromagnetic materials, AHE is typically associated with spin-orbit coupling.[98] However, in Weyl semimetals, the effect is significantly enhanced by the topological features of their electronic structure. A prominent example is the ferromagnetic Weyl semimetal $Co_3Sn_2S_2$, which exhibits an exceptionally large AHE. In this material, spin-orbit coupling drives the generation of Berry curvature, enhances scattering mechanisms, such as skew scattering and side jump, and modifies the band structure, thereby facilitating the emergence of transverse currents.[95, 99] Moreover, magnetization amplifies these effects by breaking time-reversal symmetry and inducing spin polarization, which further enhance the contributions of spin-orbit coupling to the AHE.[100] Additionally, the synergistic interplay of crystal field effects, ferromagnetism, and spin-orbit coupling, leads to the formation of Weyl points with opposite chirality (Figure 5b), marking the emergence of magnetic Weyl semimetal phase in $Co_3Sn_2S_2$. Within each bulk Brillouin zone, three pairs of Weyl points are connected by spin-filtered arcs, as shown in Figure 5c. Temperature-dependent transport and magnetization measurements of high-quality $Co_3Sn_2S_2$ reveal a ferromagnetic transition at $T_p = 175$ K (Figure 5d), confirming its magnetic Weyl semimetal nature.[95] Figure 5e presents the variation of Hall conductivity ($\sigma_H$) as a function of magnetic field at different temperatures of 100 K, 50 K, and 2

K, under the conditions where the current I//x//[$2\bar{1}\bar{1}0$] and the magnetic field B//z//[0001]. It can be observed that below 100 K, $\sigma_H$ exhibits pronounced hysteresis loops and sharp magnetic field transitions, indicating a strong AHE in $Co_3Sn_2S_2$ within this temperature range.

Similarly, in $Mn_3Sn$, an antiferromagnetic Weyl semimetal, AHE emerges despite the absence of net magnetization.[46] To investigate the spin characteristics of $Mn_3Sn$ film, Wang et al. examined its Hall effect at various temperatures, and observed pronounced AHE across the entire temperature range from 50 to 300 K. [96] Interestingly, within the temperature range of 150 to 250 K, the topological Hall effect emerges, characterized by non-flat Hall resistivity, alongside a sign reversal of the AHE at 210 K (Figure 5f left panel). Notably, the topological Hall effect reaches its maximum value at 210 K, coinciding with the magnetization peak observed at the same temperature (Figure 5f right panel). This behaviour is rarely observed in antiferromagnets, where time-reversal symmetry typically cancels out the Hall response. However, in $Mn_3Sn$, the noncollinear spin arrangement in kagome lattice structure locally breaks time-reversal symmetry, generating a net Berry curvature that drives the AHE. Additionally, it has been found that the orientation of the magnetic axis strongly affects the magnitude of the AHE, offering a tunable parameter to the transport properties of this material. Meanwhile, in $TaIrTe_4$, AHE is driven by the tilted Weyl cones, which enhance the intrinsic Berry curvature, particularly near the Fermi level, resulting in a highly anisotropic electronic dispersion.[97] Additionally, its Hall response is further amplified by the non-centrosymmetric crystal structure. While surface Fermi arcs may contribute to the Hall effect in $TaIrTe_4$, their exact role remains an open area of research.

Another interesting characteristic of Weyl semimetals is their ability to exhibit higher-order Hall responses, namely NLHE.[97] This phenomenon is driven by the Berry curvature dipole, a parameter representing a higher-order moment of the Berry curvature distribution in momentum space. NLHE emerges in materials that break spatial-inversion symmetry, enabling the generation of a transverse voltage without an external magnetic field. For example, Wang et al. observed a pronounced third-order nonlinear Hall effect in Weyl semimetal $TaIrTe_4$ at room temperature.[97]

The third harmonic of transverse voltage $V_{\perp}^{3\omega}$ exhibits a cubic relationship with the longitudinal voltage $V_{\parallel}$ across various crystal orientation angles (Figure 5g).[97] Besides third-order response, second-order Hall effect has also been observed in $TaIrTe_4$, manifesting as a Hall-like voltage or current proportional to the square of the applied electric field. Specifically, Kumar et al. observed a distinct second-harmonic transverse voltage $V_{\hat{b}-\hat{a}\hat{a}}^{2\omega}$, which is significantly larger than the longitudinal response $V_{\hat{a}-\hat{a}\hat{a}}^{2\omega}$ (Figure 5h). This transverse voltage scales quadratically with the applied current $I^{\omega}$, providing strong evidence for the second-order nature of the NLHE in $T_d$-$TaIrTe_4$.[94] It is noteworthy that NLHE is unique to systems with complex topological properties, directly reflecting the underlying asymmetry in their crystal structure.

In addition to the AHE and NLHE, Weyl semimetals like $Co_3Sn_2S_2$ are predicted to exhibit the quantum anomalous Hall effect (QAHE) under specific conditions.[99] QAHE is a quantized version of the AHE, where the Hall conductivity is quantized in units of the conductance quantum. This quantization may occur in the presence of topologically protected chiral edge states[42] and facilitate the realization of dissipationless edge conduction, a key property for the development of quantum technology devices. However, QAHE has not yet been experimentally observed in $Co_3Sn_2S_2$, where it is expected to emerge only under very low temperatures or in thin-film configurations.

In summary, the underlying physics of these Hall effects is deeply linked with the symmetry properties of the Weyl semimetals. For example, the breaking of time-reversal symmetry, observed in $Co_3Sn_2S_2$ and $Mn_3Sn$, is essential for the emergence of AHE, while inversion symmetry breaking, as in $TaIrTe_4$, is necessary for NLHE. In both cases, Berry curvature, driven by the presence of Weyl nodes, serves as the critical quantity that governs the transport behavior. Additionally, the unique surface states of Weyl semimetals can also play a role in the Hall transport properties. In systems such as $Co_3Sn_2S_2$ and $TaIrTe_4$, these surface states contribute to AHE and QAHE, particularly in thin films where surface transport becomes more prominent. To conclude this sub-sections, the required symmetry and experimental conditions for observing various Hall effects are summarized in Table 2, together with their potential device applications.

**Table 2. Hall effects in Weyl semimetals**

| Hall effect | Required symmetry (Origin mechanism) | Experimental conditions | Potential device applications | Reference |
|---|---|---|---|---|
| Ordinary Hall Effect (OHE) | None | External magnetic field | Magnetic field, current sensor | 101 |
| Anomalous Hall Effect (AHE) | Broken time-reversal (spin-orbit coupling) | Zero external magnetic field | Magnetic memory | 98, 99 |
| Topological Hall Effect (THE) | Broken time-reversal (spin structures) | Low temperature, magnetic field | High-precision magnetic sensor | 46, 58 |
| Quantum Anomalous Hall Effect (QAHE) | Conserved spatial-inversion, broken time-reversal | Low temperature, zero external field | Low-power, quantum computing | 21, 99 |
| Nonlinear Hall Effect (NLHE) | Conserved time-reversal, broken spatial-inversion | Zero magnetic field, external AC electric field | Nonlinear signal processing | 94, 97 |

### 4.4. Ultrahigh transport lifetime

Thank to the topologically protected states, Weyl semimetals not only exibit ultrahigh carrier mobility, but also an exceptionally high transport lifetime, relative to the quantum lifetime.[102] For instance, the quantum lifetime characterizes the duration over which an electron maintains its phase coherence within a particular quantum state, while the transport lifetime quantifies the average time an electron travels before experiencing a significant change in momentum direction. In conventional materials, the ratio of transport to quantum lifetimes typically remains within a few-fold range. However, in Weyl semimetals, this ratio has been theoretically predicted to approach $10^6$.[82-84] More intriguingly, the transport lifetime is predicted to exhibit negligible dependence on impurity concentration, in stark contrast to the quantum lifetime, which decreases rapidly with increasing disorder. According to the transport model, the exceptionally large ratio of transport to quantum lifetime arises from the suppression of backscattering in Weyl electrons, a phenomenon enabled by their intrinsic chirality. Specifically, when a Weyl electron encounters an impurity, it undergoes a chirality-protected lateral shift, deflecting transversely rather than reversing its momentum. This mechanism effectively preserves the electron's forward motion, thereby enhancing the transport lifetime.

Not only been theoretically predicted, the ultrahigh transport lifetime of Weyl semimetals has also been observed experimentally. For example, Cheng-Long et al. investigated Hall transport in TaAs and extracted a transport lifetime as high as 26 ps, approximately 79 times greater than the quantum lifetime of 0.34 ps.[83] Interestingly, this ratio significantly decreases as the Fermi level moves away from the Weyl point, emphasizing the critical role of Weyl electrons in enabling transport roburstness. Similarly, Tian et al. performed Hall measurements in $Cd_3As_2$ and reported a transport lifetime reaching 210 ps.[84] Together with a remarkably short quantum lifetime of 0.03 ps, the resulting ratio between transport and quantum lifetimes goes up to $10^4$. This means that, on average, it would take $10^4$ scattering events to reverse the momentum of a Weyl electron, underscoring the robustness of charge transport in Weyl materials.

In summary, the extraordinary suppression of backscattering, attributed to the topological protection and chirality of Weyl fermions, is a unique hallmark of Weyl semimetals. This transport robustness positions them as promising materials for low-dissipation electronic systems and quantum information technologies.

### 4.5. Other properties

Beyond the previously discussed characteristics, Weyl semimetals exhibit additional intriguing properties influenced by strong spin-orbit coupling. For example, in solid state systems, $PtBi_2$ stands out due to its strong spin-orbit coupling, which plays a critical role in shaping its topological properties, including the formation of Weyl semimetal phase.[38] The combination of spin-orbit coupling and inversion symmetry breaking makes $PtBi_2$ an interesting system for studying the interplay between topological effects and superconductivity. Notably, the strong spin-orbit coupling also affects its low-dimensional superconducting properties, enhancing its electronic and magnetic responses and enabling the observation of phenomena such as the Berezinskii-Kosterlitz-Thouless transition in thin flakes.[38] Similarly, in $Co_3Sn_2S_2$, high-resolution ARPES measurement reveals a substantial spin-orbit-coupling-induced energy gap of up to ~55 meV along the nodal line in momentum space.[103] This gap confirms the transition from a degenerate nodal ring state to a Weyl semimetal phase. The detection of this large energy gap provides key insights into the material's unusual electronic properties, such as its large anomalous Hall conductivity and Seebeck effect.

## 5. Materials Design and Engineering

Beyond investigating the intrinsic properties of Weyl semimetals, researchers have been actively employing advanced material design and engineering strategies to fine-tune and optimize their unique characteristics. These efforts aim to not only enhance our understanding of the fundamental behaviors of Weyl semimetals, but also pave the way for novel functionalities and practical applications. The most prominent techniques are elemental doping,[14] application of external stimuli,[13] and Moiré engineering.[33, 104] Each of these methods will be discussed in greater details in the following sections.

### 5.1. Elemental doping

Elemental doping is a powerful technique for tailoring material properties due to its ability to precisely tune the electronic band structure. This method has been extensively employed across a wide range of materials, including semiconductors, nanomaterials, two-dimensional materials, and catalysts. Recently, researchers have demonstrated the potential of elemental doping to engineer

Weyl semimetals, shedding light on the principles governing their topological phase transitions. One prominent mechanism involves the introduction of magnetic impurities through foreign atom doping, which breaks the material's symmetry and induces a topological phase transition. For example, doping Fe atoms onto the surface of ZrSiTe breaks time-reversal symmetry, causing a degenerate Dirac point to split into a pair of Weyl points.[14] Further investigation into this mechanism reveals that this time-reversal symmetry breaking is induced by the local magnetic potential generated by the magnetic impurities.

A theoretical representation of the Dirac point splitting under the influence of magnetic impurity potentials is illustrated in Figures 6a-d.[11] Figure 6a shows the impurity-perturbed dispersion relation along the $k_x$ axis for a relatively weak magnetic impurity potential. A pair of Weyl nodes emerges along the $k_x$ axis, indicating that the Dirac point has split into two Weyl nodes due to the breaking of time-reversal symmetry caused by the magnetic impurity potential.[14] In the case of a stronger magnetic impurity potential (Figure 6b), the Weyl nodes are further separated in momentum space, but localized states emerge near the Weyl nodes and smear them. These localized states are a result of the self-energy caused by impurity scattering, leading to a more complex dispersion relation. Figure 6c illustrates the Berry curvature in momentum space for a weak magnetic impurity potential. Here, the circles represent the positions of the Weyl nodes, with red and green colors representing positive and negative chirality, respectively. The Berry curvature exhibits a source and a sink, corresponding to the Weyl nodes, resembling a pair of magnetic monopoles in momentum space. Figure 6d shows the Berry curvature in momentum space for a strong magnetic impurity potential, where it is significantly modified by the presence of localized states, leading to the smearing of the Weyl nodes.

In conclusion, by tuning the potential of magnetic impurities, the electronic structure of Weyl semimetals can be effectively manipulated, enabling control over phenomena, such as Weyl node splitting, the formation of localized impurity states, and the modulation of Berry curvature. Among these, the ability to tune the position of the Weyl nodes is particularly important, as the physical properties of Weyl semimetals are highly sensitive to the energy offset between the Fermi level and the Weyl nodes.[105, 106] This sensitivity is especially crucial for phenomena and applications involving the interaction of electrons with the momentum-space magnetic field, or Berry curvature, which diverges at the location of the Weyl nodes and rapidly decays away from it.

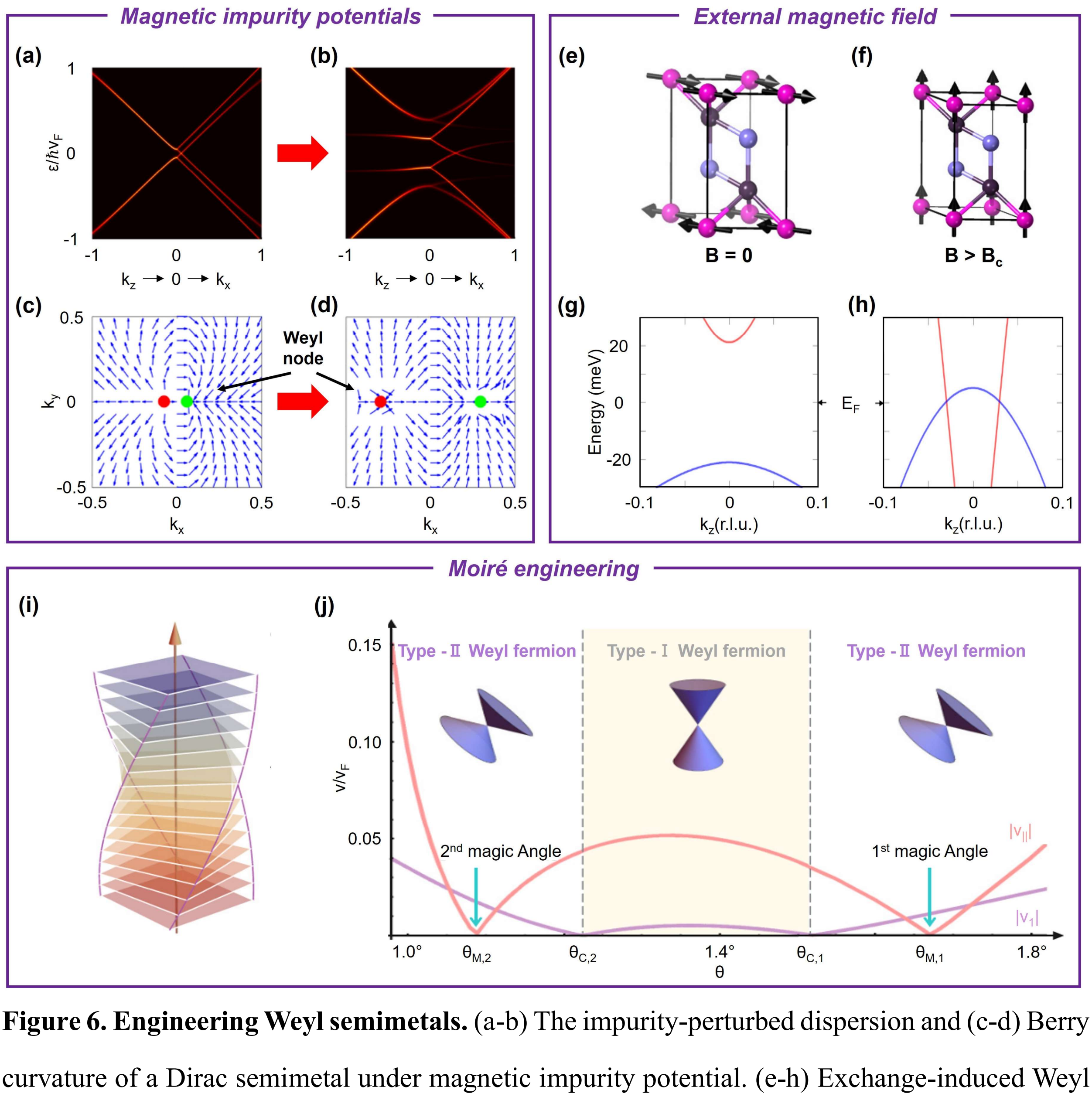

**Figure 6. Engineering Weyl semimetals.** (a-b) The impurity-perturbed dispersion and (c-d) Berry curvature of a Dirac semimetal under magnetic impurity potential. (e-h) Exchange-induced Weyl nodes in $EuCd_2As_2$. (e, g) In zero field, the Eu spins order in an A-type antiferromagnetic structure, with the spins lying in the ab plane. The corresponding band structure is gapped at Γ. (f, h) The Eu moments can be fully polarized along the c axis in a small external field $B > B_c$, lifting the double degeneracy of the bands and creating a pair of Weyl nodes. (i) In a 3D twist system, the electron

wave function is distributed along a helix due to the non-conformal symmetry of the system. (j) Two types of Weyl fermions exist in a 3D twist system when the twist angle θ varies within a certain range. Figures are reproduced with permission from (a-d) Ref.[11], Copyright 2017 American Physical Society; (e–h) Ref.[9], Copyright 2019 American Physical Society; (i-j) Ref.[107] Copyright 2020 American Physical Society, under the terms of the Creative Commons Attribution 4.0 International license.

### 5.2. External stimuli

Besides element doping, external stimuli provide an alternative approach to engineer the Weyl points. For instance, researchers have demonstrated that applying an external magnetic field to certain ferromagnetic crystals can break the time-reversal symmetry, resulting in the emergence of a Weyl semi-metallic state.[44] One notable example is the intermetallic compound $EuCd_2As_2$.[9] In the absence of an external field, $EuCd_2As_2$ exhibits antiferromagnetic ordering (Figure 6e), with a band structure showing degeneracy at the Γ point (Figure 6g). However, when an external magnetic field exceeding the critical threshold of $B_c = 1.6$ T is applied, the Eu spins fully polarize along the c-axis (Figure 6f). This polarization breaks time-reversal symmetry and lifts the degeneracy, leading to the emergence of a single pair of Weyl nodes along the Γ-A line (Figure 6h). Moreover, the position of these Weyl nodes can be dynamically tuned by manipulating the magnetization vector through an external magnetic field. This capability highlights the potential of magnetic field manipulation as a powerful tool for fine-tuning Weyl points and realizing the ideal Weyl semimetal state.

In addition to breaking time-reversal symmetry, breaking crystal spatial-inversion symmetry provide another effective approach to manipulate Weyl points and their associated topological properties. For example, HgTe-like materials are highly sensitive to mechanical strain. Applying in-plane compressive strain to these materials induces a transition to an ideal Weyl semimetal phase.[10] Specifically, this strain not only eliminates any trivial Fermi surfaces but also precisely positions the Weyl nodes at the Fermi level. Similarly, pressure can be a key tool for simultaneously tuning the topological phase and inducing superconductivity in the quasi-one-dimensional compound $(NbSe_4)_2I$.[26] By modulating the net chiral charge of Weyl points, partial disorder can be induced, leading to superconductivity in the $NbSe_4$ chains while preserving its topological properties. As a result, the strained structure becomes highly suited for exploring symmetry-protected topological features and their interactions with other phenomena, such as superconductivity, offering significant potential for both fundamental research and advanced applications.

### 5.3. Moiré engineering

In recent years, Moiré superlattices have garnered significant attention in the field of two-dimensional materials, especially in systems like graphene and transition metal dichalcogenides.[108] By twisting two atomic layers at specific angles, researchers create periodic interference patterns resembling an artificial crystal, which enables the manipulation of electronic structures and the emergence of new physical phenomena. This approach provides precise control over electronic correlation, topological properties, and band structures, all of which are governed by the interlayer twist angle.

Although Moiré superlattices have been extensively studied in two-dimensional materials, their applications to Weyl semimetals remains an emerging field, with most of research being theoretical. Similar to the observation in two-dimensional systems, the interlayer coupling in bulk topological materials could also be manipulated using Moiré superlattices, potentially leading to the emergence of Weyl semi-metallic phase.[109] One prominent study by Wu et al. showed that, in chiral twisted graphite, Weyl nodes arise with transition between type I and type II Weyl fermions depending on the twist angle (Figure 6i-j).[107] The researchers also discovered "magic angles" at which the in-plane Fermi velocity of the Weyl fermion disappears, potentially implying the emergence of superconductivity. Similarly, Yu et al. demonstrated that in the R-type $MoS_2/WSe_2$ heterostructure, the exciton spectrum exhibits Dirac and Weyl nodes. [104] Additionally, Moiré structures can introduce flat bands and van Hove singularities, leading to strong correlation effects that drive spontaneous symmetry breaking, thereby enabling novel topological and quantum geometric properties.[110] Moreover, constructing magnetic Moiré superlattices with materials, such as $CrI_3$, CrSBr, or $MnBi_2Te_4$ allows for the control of magnetic order, interlayer coupling, exchange interactions, and therefore topological phases.[111] Similarly, Moiré superlattices incorporating materials with broken inversion symmetry, such as ferroelectric crystals and anisotropic compounds, can induce cooperative effects, providing a novel pathway for creating Weyl semimetals.[108, 109, 111] Overall, these developments highlight the growing potential of Moiré superlattices as transformative tools for exploring and manipulating the unique properties of Weyl semimetals.

### 5.4. Other methods

In ultra-cold quantum gas systems, the tuning of spin-orbit coupling is a promising approach for realizing the ideal Weyl semimetal. One proposed method to achieve this involves the use of a three-dimensional optical Raman lattice to construct the required spin-orbit coupling, allowing for the measurement of Weyl points through advanced imaging techniques.[112] Lu et al. described a specific beam configuration to realize this structure, emphasizing the role of laser polarization and frequency components in tuning the interactions.[113] Precise optical control facilitates the formation of Weyl nodes with well-defined chirality. Their calculated topological phase diagram illustrates the regions with varying numbers of Weyl nodes, ranging from 0 to 8. Among them, the phases with only two Weyl nodes are of particular interest, since they mark the ideal Weyl semimetal phase. Through experimental observations, the Weyl nodes can be identified based on their positive and negative chirality, confirming the successful realization of the ideal Weyl semimetal. This system establishes a novel platform for investigating topological phenomena, offering significant potential for exploring fundamental particle interactions and advancing quantum technologies.

## 6. Functional Devices

Thanks to their unique properties, Weyl semimetals have garnered significant attention in application-oriented research over the past decade. These studies encompass a wide range of fields, including thermoelectrics, spintronics, photonics, and energy (Table 3). The following sub-sections will provide an in-depth exploration of these potential applications of Weyl semimetals in each of these areas. In addition, the ongoing research is also exploring their use for high-performance magnetic sensors and quantum computing platforms.[91, 114] However, these topics fall beyond the scope of this paper, and interested readers could refer to the references.[63, 115]

**Table 3. Fundamental properties of Weyl semimetals essential for various applications**

| Device Applications \ Fundamental Properties | | Surface States | High Carrier Mobility | Chiral Anomaly | Superconductivity | Exotic Hall Effect | Spin-Orbit Coupling | Light-Matter Interaction |
|---|---|---|---|---|---|---|---|---|
| Thermoelectrics | Longitudinal thermoelectric | | ✓ | | | | | |
| | Transverse thermoelectric | | ✓ | | | ✓ | | |
| Spintronics | Topological transistors | ✓ | ✓ | ✓ | | | | |
| | Superconducting spintronics | | | ✓ | ✓ | | ✓ | |
| | Chiral spin currents | ✓ | ✓ | ✓ | | | ✓ | |
| | Spin-orbit torque devices | | ✓ | | | | ✓ | |
| Photonics | Chiral, OAM detectors, emitters | | | ✓ | | | | ✓ |
| | Nonlinear optics | | | | | | | ✓ |
| Energy | Hydrogen evolution photocatalyst | ✓ | ✓ | | | | | |
| Others | Magnetic sensors, memories | | | | | ✓ | | |
| | Topological superconductor | ✓ | | | ✓ | | | |

### 6.1. Thermoelectrics

Thermoelectricity, which involves the direct conversion of temperature differences into electrical voltage and vice versa, plays a critical role in advancing sustainable energy technology. Studies have demonstrated that Weyl semimetals such as $WTe_2$, TbPtBi, and intercalated $T_d$-$MoTe_2$ exhibit remarkable thermoelectric behaviors, offering a unique platform that holds promise for both fundamental research and practical applications. Specifically, the presence of "massless" Weyl carriers enhances mobility and facilitates efficient carrier transport, allowing these materials to achieve high electrical conductivity.[6, 116, 117] Additionally, strong phonon scattering and enhanced electron-phonon coupling contribute to their low thermal conductivity. [118, 119] Moreover, the asymmetric density of states near Weyl nodes result in an enhanced Seebeck coefficient. As a result, these materials exhibit a remarkably high thermoelectric figure of merit (ZT). This makes Weyl semimetals highly promising for efficient thermoelectric energy conversion, bridging the gap between fundamental physics and sustainable energy solutions.

Figure 7a highlights a major advantage of Weyl semimetals for thermoelectric applications. Besides Seebeck effect, which occurs parallel to the temperature gradient (left panel), Weyl semimetals exhibit transverse thermoelectric properties, observed as the Nernst effect in the presence of a magnetic field (right panel). This effect is particularly noteworthy as it represents the ideal goal of solid-state cryogenic heat pumping, with the potential to reduce dependency on liquid nitrogen or helium for cryogenic cooling applications.[116] Importantly, from traditional thermoelectric point of view, such transverse effects are challenging to be achieved within a single material, and usually requires stacking of multiple layers, which increases complexity.[120] In terms of the magnitude of Nernst effect, the layered Weyl semimetal $WTe_2$ stands out due to its extraordinary Nernst power factor, which surpasses the conventional Seebeck power factors of state-of-the-art thermoelectric semiconductors by more than one order of magnitude.[61] As shown in Figure 7b, the temperature dependent Nernst power factor of $WTe_2$ reaches an outstanding value of over 3 $W/m.K^2$ at approximately 15 K, significantly outperforming conventional thermoelectric materials.[61] These remarkable properties position $WTe_2$ as a promising candidate for flexible micro-/nano-thermoelectric heat-pumping devices, especially in the cryogenic temperature regime.

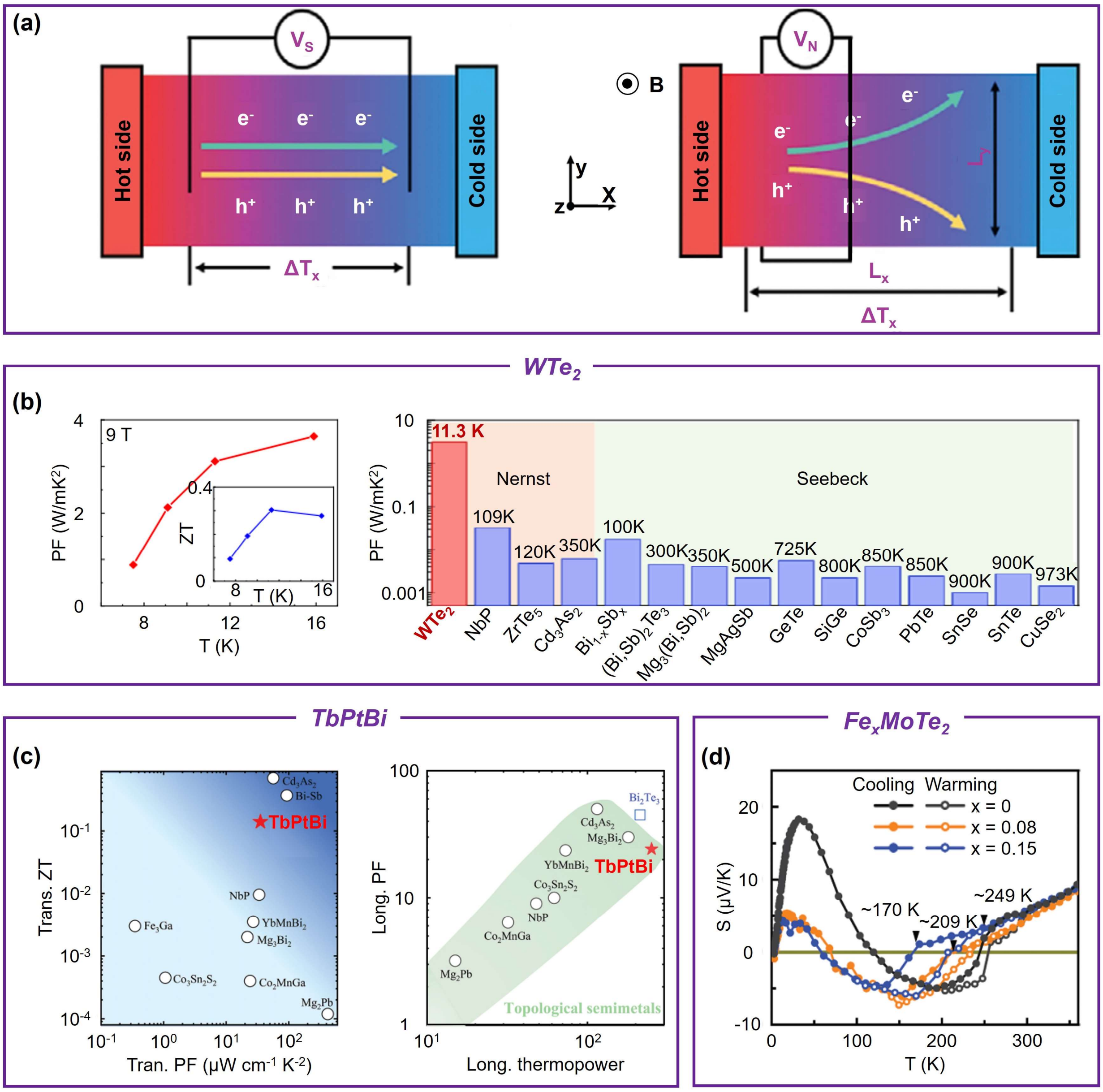

**Figure 7. Thermoelectric performance of Weyl semimetals.** (a) Schematic of thermoelectric measurement setups. (b) Left panel: Temperature dependence of Nernst power factor of $WTe_2$. Right panel: Comparison of peak power factor of $WTe_2$ with other typical thermoelectric materials. (c) Left panel: Comparison of peak transverse power factor and ZT of TbPtBi with other typical transverse thermoelectric materials. Right panel: Comparison of peak longitudinal thermopower and power factor of TbPtBi with other representative thermoelectric materials. (d) Temperature dependent Seebeck coefficient (S) for $T_d$/1T′-$Fe_xMoTe_2$ samples. The solid triangles indicate in the first sign change temperatures in S. Figures are reproduced with permission from (a) Ref.[39], Copyright 2022 Wiley; (b) Ref. [61] , Copyright 2022 Springer Nature, under a Creative Commons

Attribution 4.0 International License (CC BY 4.0); (c) Ref. [59], Copyright 2023 Wiley; (d) Ref.[8], Copyright 2023 Wiley.

Similarly, the magnetic Weyl semimetal TbPtBi exhibits both transverse and longitudinal thermopower with exceptionally high power factors, even at room temperature.[59] Specifically, at 300 K under a 13.5 T magnetic field, the transverse thermopower reaches 214 μV/K, with a power factor comparable to the most efficient thermoelectric materials available. This enhancement in thermoelectric performance, attributed to the bipolar effect and large Hall angle, is further supported by first-principles calculations, confirming the remarkable thermoelectric capabilities of TbPtBi.[59] Figure 7c left panel compares the peak transverse power factor and transverse figure of merit (transverse ZT) of TbPtBi with other representative transverse thermoelectric materials. It is clear that TbPtBi exhibits significantly higher power factor and ZT than most of these materials, except for Bi-Sb and $Cd_3As_2$. This is especially impressive given that many Weyl semimetals remain unexplored for thermoelectric applications. Figure 7c right panel compares the peak longitudinal thermopower and power factor of TbPtBi with other thermoelectric materials and $Bi_2Te_3$, demonstrating competitive performance even in traditional thermoelectric properties.

On another innovative front, Fe-intercalated $T_d$-$MoTe_2$ introduces a new tuning knob of the thermoelectric properties of Weyl semimetals. Specifically, electrical and thermoelectric transport measurements reveal a phase transition temperature that can be tuned by varying Fe concentration (Figure 7d).[8] This tunability, combined with the suppressed transition temperature and the Kondo effect arising from the coupling between conduction carriers and local magnetic moments of intercalated Fe atoms, results in a rich electronic phase diagram with opportunities for exotic electronic states exploration. It is worth noting that achieving such a highly tunable phase transition temperature is generally challenging in bulk thermoelectric materials, and often results in drastically altered band structure. At the same time, tuning phase transition temperature is crucial for optimizing both transport and mechanical properties, both of which are crucial for module-level performance, particularly in matching the performance between n-type and p-type

legs.[61]

In summary, Weyl semimetals offer a promising platform for the study and application of thermoelectric effects. The synergistic interplay of the large anomalous Nernst effect, high carrier mobility, and mechanical flexibility positions Weyl semimetals as key candidates for future thermoelectric applications. Since thermoelectric devices demand no moving parts, this research opens the door to silent, vibration-free applications in fields such as medical storage, precise temperature control for laser diodes and infrared detectors, and low-temperature cooling for quantum computing and spintronics. Furthermore, owing to their intrinsic high-performance and the ability to be shaped on-demand, Weyl semimetals could enable the development of innovative systems, such as wearable heat harvesting devices and high-sensitivity sensors for biomedical applications.[121]

### 6.2. Spintronics

The unique topologically-protected spin-polarized surface states of Weyl semimetals make them promising candidates for spintronic applications, particularly in devices where the spin and charge degrees of freedom are coupled. One of the primary research directions is the utilization of these surface states to develop novel spin current generators, spin filters, and spin-based logic devices. The key mechanism behind these devices is spin-momentum locking, where the spin orientation of an electron is tied to its momentum, enabling the generation of highly efficient spin currents. A notable contribution to this field was made by Krieger et al., who used ARPES to observe the Weyl spin-momentum locking of multifold fermions in the chiral topological semimetal PtGa.[122] The study revealed that the electron spins of the Fermi arc surface states are orthogonal to their Fermi surface contours for momenta close to the projection of the bulk multifold fermions at the Γ-point, consistent with the expected Weyl spin-momentum locking behavior. This Weyl spin-momentum locking could pave the way for the development of energy-saving memory devices and Josephson diodes based on chiral topological semimetals.[123]

In addition, the intrinsic chiral nature of the Weyl points and their associated Fermi arcs can be harnessed to generate chiral spin currents or topological spin Hall effects, where the spin current flows in a specific direction depending on the chirality of the Weyl nodes. For instance, Bainsla et al. explored the out-of-plane spin-orbit torque efficiency of the topological Weyl semimetal $TaIrTe_4$ for spintronic applications.[124] The researchers fabricated $TaIrTe_4/Ni_{80}Fe_{20}$ heterostructures and used spin-torque ferromagnetic resonance and second harmonic Hall measurements to study the spin Hall effect in this system. Their results revealed a significant out-of-plane damping-like spin-orbit torque efficiency at room temperature. The out-of-plane spin Hall conductivity was estimated to be $(4.05 \pm 0.23) \times 10^4$ $(\hbar/2e)(\Omega m)^{-1}$, which is an order of magnitude higher than that reported in other materials, highlighting the potential of $TaIrTe_4$ for energy-efficient magnetization switching in spintronic devices.

Apart from these studies, considerable research has been dedicated to exploiting the spin properties of Weyl semimetals for spintronics applications. $TaIrTe_4$ has been demonstrated to be an effective spin current source, due to its low crystal symmetry, which enhances vertical spin polarization and significantly improves spin-orbit torque efficiency.[5] Similarly, $WTe_2$, owing to the spin-momentum locking effect of its Fermi arcs, exhibited significantly enhanced spin conductivity and field-effect torque at low temperatures in $WTe_2$/Py bilayer structures. [62] Here, the spin-polarized surface states in $WTe_2$ also induce tunnel resistance switching when subjected to an applied magnetic field. Meanwhile, a $TaIrTe_4$/Ti/CoFeB heterostructure has shown the ability to switch perpendicular magnetization without the need for an external magnetic field.[5] Likewise, in $Co_2MnGa$ thin films, the topological ferromagnetic state can be electrically controlled via spin-orbit torque, with transport measurements revealing a large AHE and negative magnetoresistance.[103] These studies highlight the crucial role of spin-orbit coupling and spin-momentum locking in Weyl semimetals for the development of advanced spintronic device.

### 6.3. Photonics

The light-matter interactions in Weyl semimetals are particularly interesting due to the chirality of their Weyl nodes. For example, in materials like TaAs[125] or RhSi[48], photocurrents are generated under circularly polarized illumination, a phenomenon known as circular photogalvanic effect (CPGE). Notably, the specific circular polarization required for CPGE is determined by the chirality of the Weyl nodes, revealing a new degree of freedom in the material and providing a novel method for storing and transmitting information (Figure 8a).[68] Furthermore, the magnitude of the CPGE associated with optical transitions near the Weyl node is proportional to the node's topological charge, which takes discrete values of +1 or -1. Therefore, in chiral Weyl semimetals, the absence of degeneracy between nodes of opposite topological charge ensures the activation of CPGE, making it a distinctive feature of these systems.

In addition, TaAs has also been shown to radiate chiral terahertz wave, where the polarization of the terahertz radiation can be controlled by adjusting photocurrents generated via the CPGE effect.[125] These insights suggest new design concepts for chiral photon sources using quantum materials and open up opportunities for advancements in ultrafast optoelectronics. Apart from these optoelectronic applications, it has been demonstrated that when circularly polarized light is directed along the separation axis of Weyl nodes, the resulting chiral magnetic instability anomalously amplifies the reflected light on the surface of the Weyl semimetal, highlighting their potential as chiral light amplifiers with helicity selectivity.[126]

Thanks to their chirality, Weyl semimetals exhibit unique interactions not only with the polarization of light, but also with its orbital angular momentum (OAM). In the context of light, OAM refers to the angular momentum carried by a light beam due to its helical or twisted wavefront, where the phase of the wavefront varies in a helical manner around the beam's axis. This results in a mode of light that has a defined amount of angular momentum, and it can be characterized by an integer value (often denoted as $m$), which represents the number of twists or helices in the wavefront.[127] In communication systems, by adjusting the OAM of a light beam, optical signals can be encoded and decoded more efficiently, leading to increased transmission rates and information capacity.[128] In microscopy, OAM facilitates super-resolution imaging techniques, enabling the visualization of minute structures.[129] In quantum information, OAM is employed to encode and transmit qubits, allowing for the execution of complex quantum operations and communications.[130] Furthermore, OAM enables precise manipulation of microparticles and atoms, offering new opportunities in micro- and nanotechnology.[127]

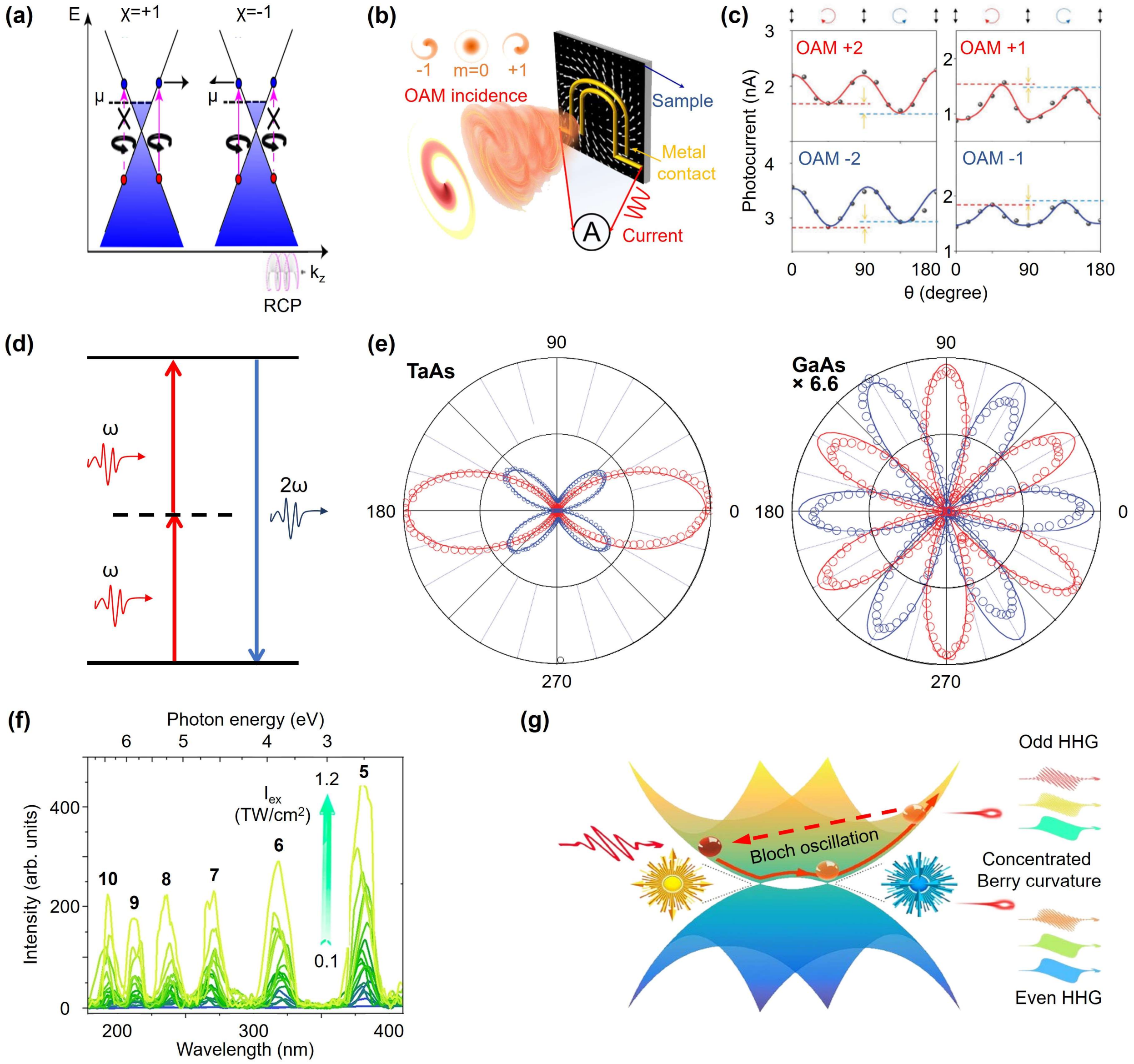

**Figure 8. Photonics based on Weyl semimetals.** (a) The chirality of Weyl semimetals makes them to be selective to polarized light. For different topological charge χ, right-handed circularly polarized (RCP) light induces transitions at different k-points. (b) The OAM detection setup utilizing Weyl semimetal. (c) Different OAM orders generate distinct currents via the orbital photogalvanic effect. (d-g) Principles of SHG (d) and HHG (g), together with giant signals (e-f)

observed in Weyl semimetals. Figures are reproduced with permission from (a) Ref.[68], Copyright 2017 Springer Nature; (b) Ref.[128], Copyright 2020 AAAS; (c) Ref.[18], Copyright 2022 Wiley; (e) Ref.[35], Copyright 2016 Springer Nature; (f-g) Ref.[54], Copyright 2021 Springer Nature, under a Creative Commons Attribution 4.0 International License (CC BY 4.0).

For various OAM-related applications, it is essential to directly detect the OAM of light in the mid-infrared range. With a photodetector based on the Weyl semimetal $TaIrTe_4$, Lai et al. successfully demonstrate a broadband response from visible light to mid-infrared.[18] The photodetector has been designed with a unique electrode configuration to directly detect the OAM orders, through the current generated via orbital photogalvanic effect (OPGE). OPGE is related to the orbital motion of electrons in response to linearly polarized light or an oscillating electric field. In OPGE, light drives the charge carriers into an asymmetric distribution due to their orbital motion, leading to a photocurrent.[18] The schematic diagram (Figure 8b) illustrates the measurement setup for OAM detection using a U-shaped electrode device. The measured photocurrent amplitude under Laguerre-Gaussian beam excitation, with OAM values ranging from +2, +1 (red curve) to -1, -2 (blue curve), is shown in Figure 8c. It is evident that as the OAM order of light switches from $+|m|$ to $-|m|$, the generated radial current reverses its sign, and its amplitude is proportional to the OAM order (m), clearly confirming the detection of the OAM order of light. This characteristic of OAM detection could be potentially integrated with the nanophotonic metasurfaces with the two-dimensional arrangement of nanostructures to achieve the resonant light-matter interactions, especially the metasurface designs capable of engineering chirality and OAM. [131-137]

Certain Weyl semimetals, particularly those in the transition metal monopnictide family (e.g., TaAs and NbAs), exhibit highly anisotropic nonlinear optical responses. This behavior arises from their non-centrosymmetric crystal structure, combined with a magnetization contribution associated with the concentrated Berry curvature near Weyl nodes. As polar semi-metals, these materials demonstrate significant second-order optical polarization, leading to pronounced effects, such as optical rectification and second harmonic generation (SHG).[138] Specifically, in SHG, the medium's charges respond to the light field in a nonlinear manner, causing the material to radiate wave with twice the frequency of the incident wave. Figure 8d presents a schematic for the SHG, where two low-energy photons are converted into one photon with twice the energy. As a result, the relationship between the intensity of the generated harmonics and the incident wave is quadratic. Extensive studies on the SHG response in transition metal monopnictides have revealed exceptionally large values of second-order nonlinear susceptibility ($\chi^{(2)}$), reaching nearly one order of magnitude greater than those observed in traditional electro-optical materials. For example, figure 8e shows the SHG intensity in TaAs, which is several times higher than that of the conventional nonlinear material GaAs.[35] Furthermore, the SHG from the electric quadrupole contribution in magnetic Weyl semimetal $Co_3Sn_2S_2$ has been experimentally demonstrated.[138] Combining Weyl semimetal with plasmonic or dielectric nanophotonic cavities could potentially enhance the nonlinear SHG conversion efficiency by orders of magnitude.[139-147]

Beyond SHG, its multi-photon counterpart, higher harmonic generation (HHG), has also been investigated in Weyl semimetals.[54] Figure 8f plots the intensity of various HHG signals in β-$WP_2$ crystal at different incident laser intensities, ranging from 0.1 to 1.2 TW/cm$^2$.[54] This confirms that, like SHG, HHG signals emerge at integer multiples of the incident beam frequency, with their intensity scaling as an integer power of the incident intensity. The underlying mechanism of HHG can be attributed to intraband Bloch oscillations (Figure 8g). Specifically, odd-order HHG arises from the nonlinear acceleration of electrons under an intense laser electric field, whereas even-order HHG requires an additional "momentum-space Lorentz force" coming from the concentrated Berry curvature near Weyl points.[148] These accelerated electrons subsequently emit high-energy photons, which are observed as higher-order harmonics. Due to the periodic nature of the light field, these harmonics appear as integer multiples of the incident laser frequency, often extending into the extreme ultraviolet or X-ray regions.[149]

## 6.4. Energy

The unique electronic properties of Weyl semimetals, particularly their high carrier mobility and robust topological surface states, can be harnessed to significantly enhance photocatalytic hydrogen evolution reaction (HER). Here, photocatalytic HER refers to a process in which light energy drives a chemical reaction that splits water molecules into hydrogen and oxygen gas using a photocatalyst, offering a clean and sustainable method to produce hydrogen fuel.[150] Its stability can be enhanced by improving charge carrier lifetime and ensuring catalyst durability.[64] As Weyl semimetals provide superior electron transport paths through their topological surface states, they can accelerate charge transfer, thereby increasing HER rates. This feature surpasses traditional local catalytic site enhancement strategies in transition metal dichalcogenides, offering a novel approach to catalyst design.[151] Additionally, the band gap inversion effect in Weyl semimetals results in a high-density state distribution near the Fermi level, facilitating the separation of photogenerated electron-hole pairs and further enhancing photocatalytic performance.

Experimental studies have demonstrated the high efficiency of Weyl semimetals and nodal line semimetals in HER catalysis. For example, $Cu_2Si$ and $Cu_2C_2N_4$ exhibit catalytic performance comparable to platinum, offering a cost-effective alternative to noble metal catalysts.[152] Besides cost considerations, Weyl semimetals, such as $NbIrTe_4$ and $TaIrTe_4$, exhibit excellent HER rates of 18,000 and 14,000 μmol/g, respectively, under visible light irradiation. In comparison to traditional photocatalysts, these Weyl semimetals exhibit a catalytic performance that is 10 to 100 times higher, particularly in photocatalytic hydrogen production, where their efficiency can reach mmol/g.h range, compared to μmol/g.h range of conventional materials like $TiO_2$.[150, 152]. This enhancement is attributed to topology-induced band inversion in M-$IrTe_4$, which increases the density of M d-states near the Fermi level, driving more efficient HER than conventional materials.[153] Furthermore, topological Weyl semimetals like NbP, TaP, NbAs, and TaAs outperform conventional transition metal disulfides in catalytic performance due to the robust topological surface states and large room-temperature carrier mobility generated by bulk Dirac bands.[23-25] This goes beyond conventional strategies that focused on increasing the number and activity of local catalytic sites, paving a new way for the discovery of novel catalysts from the emerging field of topological materials.[151]

Overall, Weyl semimetals have emerged as promising candidates for various applications, including thermoelectrics, spintronics, photonics, and energy. Their exceptional properties, such as high carrier mobility, robust surface states, and strong light-matter interactions, enable functionalities beyond those of conventional materials. These unique characteristics position Weyl semimetals at the forefront of advanced materials research, with the potential to drive next-generation technologies. To conclude this section, key Weyl semimetals and their current research trends are summarized in Table 4, providing insights into their growing impact across multiple fields.

**Table 4. Weyl semimetals for functional devices**

| Field | Weyl semimetals | Research trend | Reference |
|---|---|---|---|
| Thermo-electrics | NbP | Nernst thermoelectric power factor | 39 |
| | TbPtBi | Transverse and longitudinal thermopower | 59 |
| | $WTe_2$ | Ultrahigh transverse power factor | 61 |
| | GdPtBi | Chiral anomaly induced thermopower | 86 |
| | $YbMnBi_2$ | Anisotropy boosts transverse thermoelectric | 154 |
| Spintronics | $TaIrTe_4$ | Quantum spin Hall insulator | 21 |
| | $Co_3Sn_2S_2$ | Spin-orbit coupling effect | 103 |
| | PtGa | Weyl spin-momentum locking | 122 |
| | $TaIrTe_4/Ni_{80}Fe_{20}$ | Large out-of-plane spin-orbit torque | 124 |
| | 2D Weyl semimetal | Spin filter transistor | 155 |
| Photonics | $TaIrTe_4$ | Mid-infrared OAM detection | 18 |
| | $NbIrTe_4$ | Room temperature THz topological response | 53 |
| | $WTe_2$ | Photocurrent detection of OAM | 128 |
| | TaAs | Chiral THz wave emission | 125 |
| | $TaAs/Co_3Sn_2S_2$ | Second-harmonic generation | [138, 156] |
| | $WP_2$/ $MoS_2$ | High-harmonic generation | [54, 148] |
| Energy | $MoTe_2$, TaS, NbAs, TaAs, NbP, TaP | Hydrogen evolution catalysts | [151] |

## 7. Summary and Outlook

In summary, this review began by introducing Weyl semimetals through their fundamental concepts and features, such as the presence of Weyl nodes, hosting of "massless" carriers, and topological surface states, offering a solid background for researchers exploring this emerging field. Their unique fundamental properties, including chiral anomaly, superconductivity, various Hall effects, and ultrahigh transport lifetime, were then reviewed to establish a foundation for discussing a range of potential applications. Subsequently, the engineering methods for optimizing these properties, such as element doping, external stimuli, and Moiré engineering, were presented, providing insights into tailoring these materials for specific applications. Lastly, the current utilization of Weyl semimetals was examined, emphasizing their advantages and unique contributions to various advanced technologies, including thermoelectrics, spintronics, photonics, and energy. Before concluding the review, we would like to highlight current challenges and future outlook for Weyl semimetals.

### 7.1. Challenges

Despite the significant potential of Weyl semimetals, several practical challenges hinder the widespread commercialization of their functional devices. Substantial advancements are needed to fully exploit their unique physical properties and transition them from laboratory demonstrations to scalable real-world applications. One key challenge lies in the non-ideal band structures of existing Weyl semimetals, where multiple trivial bulk and surface states coexist with topological states at the Fermi level. These non-topological states obscure hallmark features of Weyl fermions, such as the chiral anomaly and Fermi arc surface states, by introducing extraneous electronic responses that dominate the transport behavior. Consequently, researchers have to be cautious when attributing observed electrical properties, such as superconductivity, to Weyl electrons in non-ideal Weyl semimetals. This distinction is especially critical for applications requiring the coexistence of multiple quantum phenomena, such as topological superconductivity. Moreover, topologically trivial states introduce additional scattering channels that degrade the coherence of Weyl fermions and enable hybridization with surface Fermi arcs, thereby compromising their topological protection.[157] Therefore, the realization and utilization of an ideal Weyl semimetal with a clean electronic structure is a crucial prerequisite for advancing the development of practical device applications.

Another major challenge is the extreme sensitivity of Weyl semimetal properties to the position of the Fermi level relative to the Weyl points. This sensitivity is particularly critical for phenomena that depend on the interaction of electrons with the momentum-space magnetic field, or Berry curvature. Since the Weyl point acts as a monopole in momentum space, the Berry curvature diverges at the Weyl point and rapidly decays away from it. As a result, the magnitude of unique topological effects, such as the chiral anomaly, significantly diminishes when the Fermi level deviates from the Weyl point.[105, 106] Moreover, displacement of the Fermi level disrupts the surface states arising from Fermi arcs connecting Weyl points, which adversely impacts device applications that rely on them, such as topological transistors and photocatalysts. To address this issue, it is essential to precisely tune the Fermi level, which can be achieved through various discussed material engineering strategies, including elemental doping and Moiré engineering.

Furthermore, many topological phenomena observed in Weyl semimetals, such as QAHE and chiral-anomaly-related transport, are currently prominent only at very low temperatures. For practical applications, it is crucial to develop materials and engineering strategies that enable these effects to be realized and utilized at room temperature. Enhancing their stability under ambient conditions remains a key challenge that must be overcome to fully harness the technological potential of Weyl semimetals.

Additionally, some of the most intriguing phenomena in Weyl semimetals, such as the chiral anomaly, require the simultaneous application of strong electric and magnetic fields. This requirement introduces significant complexity for practical device integration, as it necessitates compact and scalable field generation mechanisms. To enhance the usability of Weyl semimetals in real-world applications, it is crucial to develop innovative approaches that either replicate these conditions more feasibly or identify alternative mechanisms that do not rely on such stringent requirements.

Regarding thermoelectric properties, while certain Weyl semimetals exhibit a giant Nernst effect, this phenomenon is typically pronounced only within a narrow temperature range, limiting its practical utility. Moreover, the magnetic fields required to induce large Seebeck or Nernst effects in nonmagnetic Weyl semimetals are generally very high (typically exceed 5 T). To overcome these limitations and realize cryogenic cooling in transverse thermoelectric devices, it is essential to develop a set of materials with complementary operating temperature ranges and requiring low external magnetic fields (less than 1 T). These materials must maintain high performance across a broad temperature range, from room temperature down to cryogenic conditions, such as those achievable with liquid nitrogen or liquid helium.

Achieving wafer-scale, uniform synthesis of Weyl semimetals, especially in thin film configuration, while maintaining the high purity and structural consistency required for optimal device performance remains a great challenge. Currently, non-scalable advanced growth techniques, such as MBE, are necessary for ensuring reproducible production with relatively well-controlled material properties.[158, 159] Nevertheless, the carrier mobility of grown Weyl semimetal thin films remains lower than that of single crystals. On the other hand, wafer-scale techniques, such as atomic layer deposition, often result in poor sample quality.[160, 161] A promising compromise between scalability and material quality can be offered by CVD, which has been demonstrated for the growth of Weyl semimetal NbAs with a ultrahigh conductivity and large magnetoresistance.[35,

[36] However, the resulting products are generally limited to nanometer size with limited control over shapes and dimensions. To enable practical integration of Weyl semimetals into functional electronic and optoelectronic devices, substantial improvements in the growth quality across all these synthesis techniques will be essential.

Finally, integrating Weyl semimetals into functional device architecture presents significant challenges. Establishing high quality interfaces between Weyl semimetals and other materials, such as electrical contacts or substrates is essential for minimizing energy losses and maintaining their unique electronic properties.[162-164] However, many critical issues, including defects, material stability, and scalability, must be addressed to ensure reliable and efficient device performance.[29, 63, 165] For example, in traditional device fabrication methods that rely on lithography followed by metal deposition, the surface of the Weyl semimetal is often prone to physical and chemical damage. This damage can result in the formation of a Schottky barrier at the metal-semimetal interface, which significantly increases the contact resistance and hinders efficient carrier injection. Additionally, Fermi level pinning is also commonly induced due to interfacial states or defects.[166] This severely impairs the electrostatic control over the carrier density, degrading the performance of electronic and optoelectronic devices. One promising strategy to overcome these issues is laser-induced phase patterning.[167] In this approach, localized laser irradiation is used to induce phase transitions in selected regions of the material, enabling the formation of thermodynamically stable heterophase homojunctions. This technique eliminates the need for complex transfer or patterning processes, minimizes surface damage, and naturally creates low-resistance Ohmic contacts.

Beyond surface damage, lithography and focused ion beam processing may introduce defects and unintentional doping in Weyl semimetals, thereby altering their intrinsic electronic properties.[168] This is particularly critical given the sensitivity of Weyl semimetals' properties to the precise position of the Fermi level, as discussed earlier. To mitigate processing-induced damage and doping, researchers are actively exploring low-energy ion beam techniques and optimizing lithography processes.[169, 170]. Furthermore, element diffusion at the interface can lead to several issues, including degraded contact quality and unintended doping of the Weyl semimetal, whose consequences are discussed above. This diffusion can also result in chemical instability at the interface, which compromises the long-term stability and reliability of the device.[171] Over time, such instability may cause performance degradation, particularly under operating conditions where the material is exposed to high temperatures or external stress. In summary, developing device fabrication techniques that are fully compatible with Weyl semimetals remains an ongoing challenge that must be addressed to enable their widespread technological application.[172]

### 7.2. Outlook

Although numerous challenges persist, the promising properties of Weyl semimetals continue to inspire extensive research, with the potential to revolutionize technologies across various fields. On the fundamental side, the recent realization of an ideal Weyl semimetallic phase in a two-dimensional system,[173] albeit requiring substrate coupling rather than being intrinsic to the material, provides an excellent platform to explore novel low-dimensional quantum phenomena such as the parity anomaly.[174] On the application front, several exciting research directions are emerging for Weyl semimetals. One prominent area is the integration of two-dimensional Weyl semimetals into photonic devices with the structural resonances, such as localized plasmonic resonance,[175-182] dielectric Mie resonances,[131, 183, 184] bound-states-in-the-continuum,[185-188] Whispering-gallery mode,[189] Fano resonance,[190-192] and so on, to achieve the resonant light-matter interactions. The giant nonlinear and chiral optical responses observed in these ultrathin materials make them excellent platforms for integration into compact photonic structures towards achieving high-dimensional optoelectronic devices,[193, 194] as well as enabling advancements in nonlinear optics applications, such as quantum communication and super-resolution imaging.[195, 196]

In quantum computing, Weyl semimetals are promising candidates for hosting topologically protected superconducting states, which are crucial for the realization of long-sought Majorana fermions.[197-199] These exotic quasiparticles offer a potential pathway toward fault-tolerant quantum computing, representing a major breakthrough in the field.[91, 200] In particular, the robustness of Majorana zero modes could significantly enhance system stability, enabling effective quantum error correction. Consequently, the emergence of these modes at the surfaces of Weyl semimetals may play a pivotal role in advancing topological quantum computing, ultimately facilitating the development of reliable and scalable quantum systems.

Weyl semimetals hold significant promise for spintronic applications due to their unique spin-momentum locking and topologically protected surface states. The spin-momentum locking enables exceptionally large magnetoresistance up to hundred times greater than that of current state-of-the-art devices,[201] while the topologically protected states facilitate robust, dissipationless spin transport, potentially revolutionizing the development of faster and more energy-efficient spintronic devices. Such advancements could not only enhance the performance of existing technologies but also drive the creation of next-generation memory devices, sensors, and processors with superior speed, lower energy consumption, and higher integration density.

Overall, the application of Weyl semimetals across various fields remains in its early stages, presenting both challenges and opportunities for scientific and technological advancements. As research in this area progresses, we anticipate the development of innovative applications, ranging from more efficient energy storage and conversion systems to breakthroughs in quantum information processing and high-speed computing. We hope this review serves as a valuable resource for researchers by offering insightful perspectives and highlighting key areas where further exploration could drive future innovations.

## Acknowledgments

Z.D. would like to acknowledge the funding support from the Agency for Science, Technology and Research (A*STAR) under its MTC IRG (Project No. M21K2c0116 and M22K2c0088),

National Research Foundation via Grant No. NRF-CRP30-2023-0003, and SUTD Kickstarter Initiative (SKI) grant with the award No. SKI 2021_06_05. C.-W.Q. would like to acknowledge the funding support from the Competitive Research Program Award (NRF-CRP22-2019-0006 & NRF-CRP26-2021-0004) from the NRF, Prime Minister's Office, Singapore, and by a grant (A-0005947-16-00) from Advanced Research and Technology Innovation Centre (ARTIC), National University of Singapore.

**Authors' contributions**

Z.D., J.K.W.Y., W.G., and C.-W.Q. conceived the concept, designed the manuscript layout, and supervised the project. M.Z. wrote the review manuscript with the inputs from N.T.T.V, Z.D., W.Z., J. R.S., Y.L., J.W., A.S., and H.L. In addition, G.C. provides in-depth knowledge on the aspect of topological quantum materials. K.P.L. provides in-depth knowledge on the aspects of nonlinear Hall effect, twisted Weyl semimetal and spintronics. All authors analyzed the data, read, and corrected the manuscript before the submission.

## Uncategorized References